# A Thermochemical Database from High-throughput First-Principles Calculations and Its Application to Analyzing Phase Evolution in AM-fabricated IN718

Yi Wang[1*], Frederick Lia[2], Ke Wang[3], Kevin McNamara[1,2], Yanzhou Ji[1], Xiaoyu Chong[1], Shun-Li Shang[1], Zi-Kui Liu[1], Richard P. Martukanitz[4], and Long-Qing Chen[1*]

[1]*Department of Materials Science and Engineering, The Pennsylvania State University, University Park, PA 16802*

[2]*Center for Innovative Materials Processing through Direct Digital Deposition (CIMP-3D), Applied Research Laboratory, The Pennsylvania State University, University Park, PA 16802*

[3]*Materials Characterization Laboratory, Materials Research Institute, The Pennsylvania State University, University Park, PA 16802*

[4]*Materials Science and Engineering, University of Virginia, Wilsdorf Hall, 395 McCormick Road P.O. Box 400745, Charlottesville, VA 22904*

## Abstract

A comprehensive thermochemical database is constructed based on high–throughput first-principles phonon calculations of over 3000 atomic structures in Ni, Fe, and Co alloys involving a total of 26 elements including Al, B, C, Cr, Cu, Hf, La, Mn, Mo, N, Nb, O, P, Re, Ru, S, Si, Ta, Ti, V, W, Y, and Zr, providing thermochemical data largely unavailable from existing experiments. The database can be employed to predict the equilibrium phase compositions and fractions at a given temperature and an overall chemical composition directly from first-principles by minimizing the chemical potential. It is applied to the additively manufactured nickel-based IN718 superalloy to analyze the phase evolution with temperature. In particular, we successfully predicted the formation of $L1_0$-FeNi, γ'-$Ni_3$(Fe,Al), α-Cr, δ-$Ni_3$(Nb,Mo), γ"-$Ni_3$Nb, and η-$Ni_3$Ti at low temperatures, γ'-$Ni_3$Al, δ-$Ni_3$Nb, γ"-$Ni_3$Nb, α-Cr, and γ-Ni(Fe,Cr,Mo) at intermediate temperatures, and δ-$Ni_3$Nb and γ-Ni(Fe,Cr,Mo) at high temperatures in IN718. These predictions are validated by EDS mapping of compositional distributions and corresponding identifications of phase distributions. The database is expected to be a valuable source for future thermodynamic analysis and microstructure prediction of alloys involving the 26 elements.

---

[*]Corresponding Authors: Yi Wang: yuw3@psu.edu; Long-Qing Chen: lqc3@psu.edu

## 1 Introduction

With advances in both computational power and efficiency during the last two decades [1,2], density functional theory (DFT) calculations [3,4] have been routinely employed to predict energetics and properties of a material with a given structure at both 0K and finite temperatures. Here we make an attempt to build a comprehensive thermochemical database entirely based on DFT calculations. In particular, we obtained chemical potentials (often called Gibbs energies in the existing literature) of individual phases employing a quasi-harmonic phonon approach for the lattice contribution and Mermin statistics [5] for the thermal electronic contribution for nickel based superalloys by performing high-throughput first-principles calculations.

The present work was initially motivated by our desire to understand the thermodynamic equilibria among possible precipitate phases formed during heating and cooling cycles in additive manufacturing of IN718 alloys consisting of 10 elements (Ni, Cr, Fe, Nb, Mo, Ti, Al, Co, C, Cu). Due to the high-throughput nature of our calculations, we decided to drastically extend the database by considering 3281 compounds involving 26 elements, namely, Ni, Cr, Fe, Nb, Mo, Ti, Al, Co, C, Cu, B, Hf, La, Mn, N, O, P, Re, Ru, S, Si, Ta, V, W, Y, and Zr, which can potentially be widely applicable to a wide range of metallic alloys way beyond IN718. To facilitate the access to the database, we also develop a computational tool in python to predict the phase stability as a function of temperature.

The thermodynamic database is applied to the additively manufactured (AM) nickel-based IN718 superalloy to analyze the phase evolution with temperature. The considered phases include the ordered, disordered, and solution structures that can be potentially relevant to phases observed from the T-T-T (time-temperature-transformation) diagrams [6–9] of IN718. For this purpose, we combine the database and experimental observations from energy dispersive x-ray spectroscopy (EDS) measurements in transmission electron microscope (TEM). The input data to the computational tool for analyzing the presence of phases and their volume fractions are then simply the local alloy compositions from EDS.

## 2 DFT Derived Database

### 2.1 Stoichiometric phases

For stoichiometric crystalline phases, the present calculations include the L1$_2$-A$_3$M ($\gamma$'), D0$_{22}$-A$_3$M ($\gamma$''), and D0$_a$-A$_3$M ($\delta$) phases with A=(Ni, Co, Cr, and Fe) and M ($\neq$A)= Ni, Co, Cr, Fe, Al, B, C, Cr, Cu, Hf, La, Mn, Mo, N, Nb, O, P, Re, Ru, S, Si, Ta, Ti, V, W, Y, and Zr; C14-A$_2$M, A$_8$M, MC, (A,M)$_6$C, and (A,M)$_{23}$C$_6$ phases with A, M ($\neq$A)= Ni, Co, Cr, Fe, Al, B, C, Cr, Cu, Hf, La, Mn, Mo, N, Nb, O, P, Re, Ru, S, Si, Ta, Ti, V, W, Y, and Zr; All other known Ni, Co, Cr, or Fe containing stable compounds with Al, B, C, Cr, Cu, Hf, La, Mn, Mo, N, Nb, O, P, Re, Ru, S, Si, Ta, Ti, V, W, Y, and Zr obtained from the Materials Project [10], including phases such as the $\eta$ Ni$_3$(Nb,Ti), $\mu$ A$_6$M$_7$, and $\sigma$ A$_x$M$_y$-type of phases. Possible duplicate phases were excluded by automated inspection of the composition, the size of the primitive unit cell, and the space group symmetry employing the *pos2s* script (from YPHON) which was developed based on the *FINDSYM* module [11] of the ISOTROPY package [12].

### 2.2 Disordered phases

We considered both binary and ternary disordered phases at selected compositions using the special quasirandom structure (SQS) approach [13–15]. The idea of SQS is to construct a special supercell for which the atomic correlation functions mimic the physically most relevant ones of the perfectly random solid solutions. It is used extensively to study formation enthalpies, bond length distributions, density of states, etc. in solution phases [16,17]. In addition, lattice distortions can be studied with the SQS approach as well, by which the equilibrium position of each atom in a supercell may be obtained through minimizing the interatomic Hellmann-Feynman forces in DFT calculations. The considered phases include: i) the binary face-centered cubic (fcc) phases $X_xM_{1-x}$ ($\gamma$) at x=0.333, 0.5, 0.625, 0.667, 0.75, and 0.875 with X=Fe, Co, Ni, and Cr, and M= Al, C, Co, Cr, Cu, Fe, Hf, Mn, Mo, Nb, Ni, Re, Si, Ta, Ti, and W; the ternary fcc phases $X_{1/3}M_{1/3}Z_{1/3}$ and $X_{3/5}M_{1/5}Z_{1/5}$ (X=Fe, Co, Ni, and Cr, and M,Z=Al, C, Co, Cr, Cu, Fe, Hf, Mn, Mo, Nb, Ni, Re, Si, Ta, Ti, and W); the binary body-centered cubic (bcc) phases $M_{1/3}Z_{1/3}$; and the ternary bcc phases $X_{1/3}M_{1/3}Z_{1/3}$. Note that the present DFT study does not include the ternary $\sigma$ phase and other TCP phases [18] due to its complicated solution structure and computational cost.

*2.3 Dilute solution phases and antiferromagnetic states for Cr and Mn*

For the dilute solution phases, the present calculations include fcc-based phases $Ni_NM$ where N=31, 107, and 255 with M= Al, C, Co, Cr, Cu, Fe, Hf, Mn, Mo, Nb, Ni, Re, Si, Ta, Ti, and W. For the magnetic structures, we treat Co, Fe, and Ni as ferromagnetic. Antiferromagnetic states for Cr [19] and Mn [20] are treated with supercell containing 2 Cr and 58 Mn atoms, respectively.

*2.4 High-throughput quasiharmonic phonon calculations*

We automate the challenging high-throughput quasiharmonic calculations according to the procedure flowchart illustrated in Figure 1. We start with collecting structural information from the existing open internet database [10] and discard the duplicates. Then the batch job scripts are prepared that automate the job submissions, monitor numerical convergences both in the electronic step and ionic steps, and detect/correct runtime errors such as runtime over the specified limits, memory leakage, and prematurely aborted jobs. Additional convergence checks are made by examining the fitting errors of the calculated 0 K total energy vs volume data using analytical function, e.g. using the Morse function

$$\text{Eq. 1} \quad E_c(V) = a\{exp[-2(V-b)*c] - 2exp[-(V-b)*c]\} + d$$

where $E_c$ is the 0 K total energy per molar volume $V$ and $a$, $b$, $c$, and $d$ as fitting parameters. When convergence criteria are not met, we iterate the calculations by progressively modifying the input settings such as the method to estimate the Fermi energy, the scheme for block iteration, and the mixing scheme in the electronic step, etc. Based on the analytical expressions, we search for the approximate range for the equilibrium volumes to set up phonon calculations that also checks for convergence and the existence of imaginary phonon modes. Finally, we compute the thermodynamic properties while those structures with significant amount of imaginary phonon modes are ignored.

## 2.5 0 K static first-principles calculations

We employed the projector-augmented wave (PAW) method [1,2] together with the standard Perdew-Burke-Ernzerhof functional [21] as implemented in the Vienna *Ab-initio* Simulation Package (VASP, version 5.3). For the pseudopotentials, we have chosen those recommended by VASP 5.2. For the quasiharmonic approximation, we perform calculations at 7 fixed volumes with a volume interval of 6% around the 0 K equilibrium volume. To account for the lattice distortion, both lattice shape and internal atomic positions were allowed to relax and the Methfessel-Paxton order 1 method [22] was employed to determine the partial orbital occupancies.

The default energy cutoff was determined by setting the VASP keyword "PREC=A" during lattice relaxation, using the Γ centered $k$-meshes together with a sampling density of ~1000 points/atom. We found that the energy threshold of $10^{-6}$ eV is sufficient for the convergence of the electronic step, whereas it required an energy threshold of $10^{-8}$ eV for the relaxation of cell shape and atomic positions. After relaxation, the energy cutoff of 520 eV together with the tetrahedron method with Blöchl corrections [23] were employed to obtain highly accurate 0 K static energies and electronic densities of states, using the Γ centered $k$-meshes with a sampling density of ~5000 points/atom. The criterion for convergence check using the Morse function in Eq. 1 was less than or equal to $10^{-3}$ eV per atom.

## 2.6 Supercell phonon calculations

We calculate the phonon properties by building supercells based on the relaxed cell shape and atomic positions of the primitive unit cell according to the following matrix transformation

$$\text{Eq. 2} \quad \begin{Bmatrix} A \\ B \\ C \end{Bmatrix} = \begin{pmatrix} i_1 & i_2 & i_3 \\ i_4 & i_5 & i_6 \\ i_7 & i_8 & i_9 \end{pmatrix} \begin{Bmatrix} a \\ b \\ c \end{Bmatrix}$$

where $a$, $b$, and $c$ are the lattice vectors of the primitive unit cell, and $A$, $B$, and $C$ represent the lattice vectors of the supercell. The values of $i_k$ ($k$=1-9) in Eq. 2 were determined by the criteria that the supercell contained ~128 atoms with a smallest possible aspect ratio.

We compute the interatomic force constants using the VASP keyword "PREC=A" together with a $3 \times 3 \times 3$ Γ centered $k$-meshes. The thermodynamic properties were calculated using the YPHON package [24–27].

## 2.7 Chemical potentials of individual phases

The chemical potential of a compound or a solution phase $\alpha$ is the same as its molar Gibbs free energy or Gibbs free energy per atom [28]. If we adopt the unit of energy per atom, the chemical potential of $\alpha$ phase under zero pressure is reduced to the Helmholtz free energy per atom $F$ [29–34],

Eq. 3 $\quad \mu^\alpha(T) \approx F^\alpha(T,V) = E_c^\alpha(V) + F_{vib}^\alpha(T,V) + F_{el}^\alpha(T,V) - TS_{conf}^\alpha$

where $T$ is temperature, $V$ is the averaged atomic volume, and $E_c^\alpha$ is the 0 K static energy per atom of $\alpha$ phase. In Eq 3, $F_{vib}^\alpha$ is the vibrational free energy per atom of $\alpha$ phase, which can be obtained by either the Debye model [35–37] for rough estimates, or phonon calculations for more accurate results, and $F_{el}^\alpha$ is the contribution of thermal electrons to the total free energy of $\alpha$ phase which we computed using Mermin's finite temperature density functional theory [5,29,30,38]. It should be noted that for superalloys, the electronic contributions can be significant at high temperatures; For instance, it was found that the electronic contribution could account more than 10% of the total heat capacity at 1000 K for Ni [29]. The configurational mixing entropy $S_{conf}^\alpha$ in Eq. 3 is set to zero for a completely ordered structure while for a disordered solution, we approximate it using the ideal entropy of mixing, i.e.

Eq. 4 $\quad S_{conf}^\alpha = -k_B \sum_i^n x_i^\alpha \ln x_i^\alpha$

where $n$ represents the total number of components, and $x_i^\alpha$ is the atomic fraction of component $i$ in $\alpha$ phase

We first calculate all temperature-dependent thermodynamic properties at the seven fixed volumes based on Eq. 3 [25,27,29,30], and then use spline interpolation to obtain the thermodynamic properties as a function of volume.

3    *Application of Database to Analyzing Phase Evolution*

Inconel 718 (IN718) is a nickel-chromium-iron based superalloy [8] which exhibits excellent strength, creep resistance, and high-temperature corrosion/oxidation resistance and thus has wide applications such as gas turbine blades and nuclear reactors [39,40]. Due to its relatively good weldability, it has recently been explored for AM [40]. The typical mass compositions of IN718 are 50-55%Ni, 17-21%Cr, ~17%Fe, 2.8-3.3% Mo. 4.8-5.5% (Nb and Ta), (<=1%) (Co, Mn, Al, Ti, Cu, Si, C, B, S, and P) [41,42].

In its solid state, IN718 alloy is generally a multiphase mixture of FCC (face centered cubic) solid solution gamma ($\gamma$) matrix with a number of precipitated phases, including $\gamma'$ (L1$_2$), $\gamma''$ (D0$_{22}$), $\delta$ (D0$_a$), carbides etc [43] with $\gamma'$(Ni$_3$(Al,Ti)) and $\gamma''$ (Ni$_3$Nb) acting as the major strengthening phases. Long-time exposure of the alloy to high temperature leads to the transformation of the metastable $\gamma''$ to the stable $\delta$ phase with the D0$_a$ structure and precipitation of $\alpha$-Cr [9]. The progressive formation of carbides at grain boundaries may result in a gradual transition in the fracture mode of the alloy from trans-granular to inter-granular [44].

3.1   *EDS Data: Sample preparation and TEM/EDS measurements*

The specimens came from prior research with the experimental procedures described in the work of Promoppatum et al. [45]. The experiments were conducted with an EOS M280 powder-bed fusion system with the process parameters shown in Table 1. Inconel 718 powder with an average powder size of 32.2 μm was supplied by EOS. The deposited part was removed from the build plate by electrical discharge machining (EDM) and sliced. The specimens were then hot mounted in epoxy resin, ground through a sequence of grit papers, and sequentially polished utilizing a 3-micron diamond suspension, a 1-micron diamond suspension, and a colloidal silica on a Struers Pedomax-2. The specimens were rinsed in alcohol and deionized water and examined using EDS in the as-polished condition.

The transmission electron microscope (TEM) samples were prepared using an FEI Helios 660 focused ion beam (FIB) system. A thick protective amorphous carbon layer was deposited over the region of interest, and Ga+ ions (30kV then stepped down to 1kV to avoid ion beam damage to the sample surface) were used in the FIB to thin the samples to electron transparency for TEM imaging. The specimens were imaged in TEM directly after FIB sample preparation. Scanning transmission electron microscopy (STEM) was performed at 200 kV in a FEI Talos F200X S/TEM, and EDS elemental maps of the sample surface were collected by using a SuperX EDS system under STEM mode with four detectors surrounding the sample. All the STEM images were captured by using a high angle annular dark field (HAADF) detector (Fischione) for Z-contrast imaging. The EDS elemental mapping in STEM mode was performed at 200 kV with acquisition times up to 10 minutes.

*3.2 Elemental compositional data from EDS map*

The measurements were made at four selected positions at the specimen with resolutions at the scales of 1 μm, 500 nm, 200 nm, and 40 nm, and the four HAADF images are provided in Figure 2. The HAADF images for the measurements at 1 μm, 500 nm, 200 nm show the typical coarse-columnar structures observed in the previous works [46–48] with the internal primary dendritic arms growing parallel to the deposition direction or Z-orientation [49]. For each measurement, the elemental distribution data were collected for the 9 element C, Al, Co, Cr, Fe, Mo, Nb, Ni, and Ti, digitalized with image of 512×512 pixels. Figure 3 exemplifies the EDS map for the measurement at 200 nm scale (the EDS maps for the measurements at 1 μm, 500 nm, and 40 scales were given in Figure S1, Figure S2, and Figure S3 of the supporting materials, respectively).

To extract the elemental compositional distribution data, the pixel values of an EDS map are normalized to the overall composition of IN718 with the following equation:

Eq. 5 $\quad C_i(x,y) = \frac{w_i/M_i}{\sum_{j=1}^{N_{el}} w_j/M_j} \frac{N_X N_Y pixel_i(x,y)}{\sum_{x=1,y=1}^{N_X,N_Y} pixel_i(x,y)}$

where $C_i(x,y)$ is the atomic composition for element $i$ at location $(x,y)$ on the EDS map, $N_X = 512$ and $N_Y = 512$ represent the EDS image size in pixel, $pixel_i(x,y)$ is the pixel value at $(x,y)$, $M_i$ is the atomic mass, and $w_i$ is the overall weight percentage composition of IN718: Al:0.52, C:0.021 Co:0.11, Cr:19.06, Fe:18.15, Mo:3.04, Ti:0.93, Nb:5.08, Ni 53.0 [50].

*3.3    Determining phase distributions using Dantzig simplex optimization algorithm*

To predict the spatial distributions of phases utilizing a combination of the DFT-derived thermochemical database and the EDS compositional data, we compare the chemical potential (molar Gibbs energy) of an individual phase with the so-called convex hull [51] constructed based on the chemical potentials of all phases. We then calculate the phase fractions at a given temperature and composition using the Dantzig simplex optimization algorithm [52] by performing a constrained minimization of the following objective function:

Eq. 6 $\quad \mu(T,x,y) = \sum_{\alpha=1}^{m} f^\alpha(T,x,y) \mu^\alpha(T)$

where $T$ is temperature, $\mu^\alpha$ is the chemical potential of phase $a$, $m$ is the number of phases, and $f^\alpha$ is the phase fraction of $\alpha$ at the position $(x, y)$. We minimize the local chemical potential $\mu(T)$ with respect to $f^\alpha$ subject to the following conditions:

Eq. 7 $\quad C_i(T, x, y) = \sum_{\alpha=1}^{m} f^\alpha(T, x, y) x_i^\alpha$

Eq. 8 $\quad \sum_{\alpha=1}^{m} f^\alpha(T, x, y) = 1$

where $x_i^\alpha$ is the elemental composition of phase $\alpha$. The set of $f^\alpha(T, x, y)$ that minimizes the local $\mu(T, x, y)$ represents the spatial distributions of phases at a given temperature $T$. The overall phase fraction ($\bar{f}^\alpha$) of a phase at a given temperature can be calculated as:

Eq. 9 $\quad \bar{f}^\alpha(T) = \frac{1}{N_X N_Y} \sum_{x=1, y=1}^{N_X, N_Y} f^\alpha(T, x, y)$

The above procedure has been implemented in Python 3 using the Anaconda numerical software distribution, which can directly read pixel values from graphical TEM elemental distribution maps.

### 3.4 Treatment of grain boundary effects

While the intragranular plate-shaped coherent γ" and spherical γ' precipitates are beneficial to the mechanical properties of IN718 [39,40], other precipitates are detrimental. For example, the brittle Laves phase that appears between dendrites from solute segregation [53–56] can negatively affect the high-temperature fatigue behavior of the AM-fabricated IN718 sample [54,57]. The semi-coherent δ phase particles that also appear preferentially between dendrites and at grain boundaries [58,59] may act as possible crack initiation sites during high-temperature creep [59]. The transformation of the intragranular γ" into the thermodynamically more stable δ reduces the yield strength of IN718 [59]. To examine the effect due to this transformation, we consider two scenarios: i) with or ii) without imposing a grain boundary constraint. The constraint is only applied to the δ, η, C14, $Fe_7Mo_6$, $M_{23}C$, $M_6C$, $AlNb_2$, $AlNi_2$, and $NbFe_2$ phases by confining them at grain boundaries through imposing chemical potential penalty at the locations away from the grain boundaries. The grain boundaries are located by scanning through the HAADF images for the sharp changes in the pixel values. The grain boundary distributions for the measurement at 200 nm scale are illustrated in Figure 4.

## 4 DFT-derived thermodynamics and phase evolution of AM-fabricated IN718

We predicted the phase fractions and their temperature dependence for the 1 μm, 500 nm, 200 nm, and 40 nm EDS results using the DFT-derived database and they show similar results in terms of overall phase fraction as a function of temperature. In the following, we will mainly discuss the predicted results based on the 200 nm EDS data. It is predicted that the major phases (with phase fractions >1%) in the AM-fabricated IN718 are γ, γ', γ", δ, α-Cr, and C14 phases and the $L1_0$-FeNi and η-$Ni_3$Ti phases. In principle, the γ phase in IN718 is a solid solution of Ni with 9 alloying elements. However, it is currently not realistic to deal with a 10-component random solid solution using DFT. We approximate the γ phase as a mechanical mixture of binary and ternary solutions through the special quasi-random structure (SQS) model. In particular, we obtain the temperature dependence of the γ phase through the phase fractions of the SQS structures of $FeNi_2$, $CrNi_2$, FeNi, $FeNi_3$, $Fe_3Ni_5$, CrFeNi, $FeMoNi_3$, and $CrMoNi_3$ as listed in Table 3 and Table 4.

*4.1 Predicted phase evolution*

The list of phases formed up to 1500 K predicted from the 200 nm EDS compositional data are collected in Table 2 while the temperature dependencies of the major phases (with phase fractions higher than 1%) are plotted in Figure 5(a). The temperature dependence of phase stabilities can be roughly grouped into three temperature ranges. Ordered phases are dominant below 680 K whereas the disordered γ phase is dominant at temperature greater than 1140 K with the Order-disordering phase transition temperature ranging 680 to 1140 K. The predicted phase fractions at temperature 600, 800, 1000, 1200, and 1400 K are collected in Table 3.

For comparison, we also performed thermodynamic calculations (courtesy of Carpenter Inc.) using Thermo-Calc software [60] and the commercial TCNI9 database [61]. In TCNI9, the fcc and $L1_2$ structures are grouped together as the FCC-L12 phase, and it does not include the $L1_0$-FeNi phase. For the purpose of comparison with TCNI9, the DFT predicted phase fractions for the γ'-$FeNi_3$, γ'-$AlNi_3$, and $L1_0$-FeNi phases are also combined in Figure 5(a). Both DFT and TCNI9 show two characteristic transition temperatures at ~680 K and ~1310 K. However, according to DFT database, the 680 K transition is due to the complete disordering of $L1_0$-FeNi and partially disordering of γ'-$FeNi_3$ whereas according to TCNI9 it is due to a transition from FCC-L12#3 phase to FCC-L12#1 phase. The 1310 K transition from the DFT database describes the complete dissolution of α-Cr into the γ matrix while according to TCNI9 it is due to the phase transition into 100% γ phase. The estimated difference in phase fractions of the γ phase predicted from DFT and TCNI9 databases in the temperature range of 680-1310 K is ~20%.

The enthalpy of IN718 alloy as a function of temperature from the DFT database and the TCNI9 database is compared in Figure 6 where we also included the experimental data from Hosaeus et al. [62] and recommended data by Mills [63]. It showed that the enthalpy data from DFT agree much better with the experimental data from Hosaeus et al. [62] and Mills [63] than those from TCNI9.

The entropy as a function of temperature from DFT and TCNI9 databases is presented in Figure 7. We computed the entropy of IN718 alloys from DFT database by considering the contributions due to lattice+electronic entropy ($\bar{S}_{vib+el}$) and configurational entropy ($\bar{S}_{conf}$) [29–34] as follows:

Eq. 10 $S(T) = \bar{S}_{vib+el}(T) + \bar{S}_{conf}(T) = \sum \bar{f}^\alpha S^\alpha_{vib+el}(T,\alpha) + \sum \bar{f}^\alpha S^\alpha_{conf}(T)$

The sudden changes in entropy as a function of temperature from the DFT curve are either associated with order-disorder transitions, e.g. those around 680 K and 1310 K, or due to the fact that the $\bar{S}_{conf}$ of the γ solution phase is approximated as a linear mixture of the binary and ternary SQS structures at discrete compositions. At 684 K, the entropy from DFT coincides with that from the TCNI9 database. However, above 684 K, the entropy from the TCNI9 database is substantially higher than that from the DFT database while below 684 K, the entropy from the TCNI9 database is slightly lower than that from the DFT database. This discrepancy seems to be mostly due to the much larger entropy change associated with the order-disorder transition around 684 K from the TCNI9 database than the DFT database.

Similar to the entropy calculation, we computed the heat capacity of IN718 by considering lattice and electronic contributions ($\bar{C}_{P,vib+el}$, labeled as "DFT, w/o $C_{conf}$" in Figure 8) and configurational contribution ($C_{conf}$):

Eq. 11 $C_P(T) = \bar{C}_{P,vib+el}(T) + C_{conf}(T) = \sum f^\alpha C_{P,vib+el}(T,\alpha) + C_{conf}(T)$

where $C_{conf}$ can be obtained from $\bar{S}_{conf}$ given in Eq. 10:

Eq. 12 $C_{conf}(T) = T\frac{d\bar{S}_{conf}(T)}{dT}$

However, since the γ phase was treated as a linear mixture of the binary and ternary SQS structures at discrete compositions, which may result in discontinuous changes in entropy, we cannot directly use Eq. 12 to calculate $C_{conf}$. Instead, we employ the following two schemes to calculate $C_{conf}$. The scheme#1 (labeled as "DFT, with linear $C_{conf}$" in Figure 8) imposes a linear constraint between $\bar{S}_{conf}$ and temperature:

Eq. 13 $\bar{S}_{conf}(T) \cong aT$

So that

Eq. 14 $C_{conf}(T) = aT$

In scheme#2 (labeled as "DFT, with non-linear $C_{conf}$" in Figure 8), we calculated $C_{conf}$ as an average through

Eq. 15 $C_{conf}(T) = \frac{T}{N}\sum_{n=1}^{N}\frac{\bar{S}_{conf}(T+n\Delta T)-\bar{S}_{conf}(T-n\Delta T)}{2n\Delta T}$

where $\Delta T = 10\ K$ and $N$ have been chosen to be equal to or less than 20, whenever $T - N\Delta T \geq 0$ and $T + N\Delta T \leq 1600$.

The DFT calculated heat capacities for IN718 are compared with those using the TCNI9 database in Figure 8 together with measured data for IN718 from Hosaeus et al. [62], Basak et al. [64], Brooks et al. [65], and recommended data for IN718 by Mills [63]. Since no experimental data below room temperature is available for IN718, the measured data for Alloy-718 by Lee et al. [66] is used in Figure 8 for reference. It is seen that the DFT data obtained by scheme#1 agree well with the experimental data below 800 K, while those obtained by scheme#2 show several peaks due to order-disorder transitions. Also, the calculated heat capacity by TCNI9 for IN718 showed two much higher peaks than the experimental data in the temperature range between 400 and 600 K.

*4.2  Ordered phases at low temperature and phase stabilities of $L1_0$-FeNi, $\gamma$'-Ni$_3$Fe, and $\delta$-MoNi$_3$ phases*

The present DFT calculations for IN718 show that the main ordered phases are $L1_0$-FeNi (23%), $\gamma$'-Ni$_3$Fe (27%), $\gamma$'-Ni$_3$Al (4%), $\alpha$-Cr (21%), $\delta$-NbNi$_3$ (12%), $\delta$-MoNi$_3$ (6%), and $\eta$-Ni$_3$Ti (4%). The spatial distributions of $L1_0$-FeNi, $\gamma$'-Ni$_3$Fe, $\alpha$-Cr, and $\delta$-NbNi$_3$ phases at 600 K are displayed in Figure 9. The Fe-Ni binary phase diagram of Yang et al. [67,68]) was included in Figure 5(c) (for the detailed meaning of the symbol in Figure 5(c), see the original work by Yang et al. [67,68], which largely validates our DFT predictions.

It should be pointed out that the stabilities of the $L1_0$-FeNi and $\gamma$'-Ni$_3$Fe phases were less noted in existing literature [9,69,70] on IN718. $L1_0$-FeNi, named tetrataenite in mineralogy, is a commonly accessible mineral formed naturally in all slow-cooled meteorites [71,72]. It has a tetragonal structure with lattice parameters $a$ = 2.533 Å and $c$ = 3.582 Å, [73,74] which are almost identical to the c/a ratio of 1.414 to that of an fcc structure when a corresponding tetragonal unit cell is chosen. Laboratory synthesis of bulk $L1_0$-FeNi remains challenging due to the low atomic diffusion coefficients of Fe and Ni near the order-disorder transition temperature of ~600 K [71,72,75]. The stability of $L1_0$-FeNi had been an open question [76] (see Swartzendruber et al. [73] and Yang et al. [65,66] and references therein) until recently. For example, the earlier Fe-Ni binary phase diagram [77,78] showed that the $L1_0$-FeNi phase was metastable [75,81,82] while the $\gamma$'-Ni$_3$Fe phase was stable up to 790 K. However, more recent studies demonstrate that $L1_0$-

FeNi phase is stable below 600 K [79–82], being in agreement with our DFT prediction. In contrast, TCNI9 does not include the $L1_0$-FeNi phase. Therefore, TCNI9 predicted a decomposition into the BCC-B2#2" and FCC-L12#3 phases below 600 K at the FeNi composition. In TCNI9 database, the bcc and B2 structures have been grouped together as BCC-B2 phase because they are treated using the same sublattice model. In Figure 5(a), the "BCC-B2#1" phase by TCNI9 is mainly composed of Cr while the "BCC-B2#2" phase by TCNI9 mainly consists of Fe.

The present DFT database showed that the δ phase is a combination of $MoNi_3$ in $D0_a$ structure [83] (labeled as δ-$MoNi_3$ in the present work) and δ-$NbNi_3$. The phase fraction of δ-$MoNi_3$ started to decrease rapidly as temperature increases, from 6% at 700 K to less than 0.5% at 860 K. For comparison, the δ phase was labeled as "NI3TA-D0A" in the TCNI9 database, as shown in Figure 5(a). The TCNI9 database showed roughly the same phase fraction with decreasing δ phase from 700 K to 860 K. However, the DFT database predicted that the composition of the δ is Ni with mostly Nb and some Mo whereas according to the TCNI9 the δ phase is primarily Ni and Nb with minor Ti.

The spatial distributions of the γ', bcc-Cr (α-Cr), and δ phases and the overall phase distributions (the γ phase is plotted in cyan color) at 800 K from the DFT databases are shown in Figure 10. The ordered $L1_0$ phase vanishes at 690 K, and the amount of the γ'-$Ni_3$Fe phase decreases from 27% at 670 K to 14% at 680 K and to zero at 870 K. In fact, the absence of γ'-$Ni_3$Fe phase at 870 K is in excellent agreement with the Fe-Ni binary phase diagram [67,77] which indicated that the γ'-$Ni_3$Fe phase can be stable up to 790 K. According to our DFT database, the disordered γ phase at the temperature of 680 – 1140 K is made of disordered FeNi, $FeNi_2$, $FeNi_3$, and $Fe_3Ni_5$.

Different from the TCNI9 database, the DFT results showed that the γ'-AlNi$_3$ and δ-NbNi$_3$ phases are stable in IN718 throughout the entire temperature range of 0-1500 K. However, the present DFT calculations fail to reproduce the solvus temperature ~1273 K of δ-NbNi$_3$ from experiment [6]; δ-NbNi$_3$ is a stable phase for all temperatures below it melting temperature [84]. This might be due to the neglect of the anharmonic contributions to the vibrational free energy at high temperature in our DFT and lattice dynamic calculations, or the relatively limited number of compositions considered in the analysis of solution phases containing Nb.

The spatial distribution of the δ-NbNi$_3$ phase, the α-Cr phase, and their overall phase distributions at 1200 K, and the Cr6Fe6Ni6 SQS phase, the δ-NbNi$_3$ phase, the α-Cr phase, and their overall distributions at 1400 K are marked in Figure 11. The calculations show that the disordered γ phase mainly consists of disordered FeNi$_2$, CrNi$_2$, CrFeNi, FeMoNi$_3$, and CrMoNi$_3$. It was found that the ternary random structures are important in describing the disordered γ phase at high temperatures. For example, the Cr$_6$Fe$_6$Ni$_6$ SQS structure contributes 42% of the disordered γ phase at 1400 K.

At ~1120 K, the amount of α-Cr phase starts to decrease significantly as temperature increases due to the increase of the Cr-contained ternary solutions. The results are in excellent agreement with the observation by Jia and Gu [7] on AM-fabricated IN718 based on their successive morphological changes showing that the γ phase is mainly composed of Ni–Cr–Fe. We also showed that the α-Cr vanishes at ~1310 K which is substantially higher than 890 K according to the TCNI9 database. Our results agree with the T-T-T diagram by Oradei-Basile and Radavich [6] (Figure 5(d)) which indicates that the α-Cr phase can be stable up to 1110 K. It is interesting to notice that our DFT-predicted disappearance temperature of ~1310 K for α-Cr is the same temperature at which the δ-NbNi$_3$ phase is predicted to vanish by the TCNI9 database. Experimentally, Bi et al. [85] found that α-Cr always precipitated in the vicinity of δ phase. Wlodek and Filed observed [9] that the precipitation of α-Cr and other minor phases in alloy 718 are driven by the almost complete rejection of Cr and Mo from the matrix, leading to the growth of δ phase.

*4.3 Predicted phase evolution by DFT calculations with grain boundary constraint*

The predicted phase fractions for the individual phase are collected in Table 4 by artificially imposing preferred precipitation of certain phases at grain boundaries. The stabilities of the major phases (with phase fractions higher than 1%) as a function of temperature are summarized in Figure 5(b). Except for the γ" and δ phases, the evolutions of phase fractions with increasing temperature are quite similar to those calculated without considering grain boundary constraint discussed in the previous section. The phase stabilities can also be roughly grouped into the same three temperature stages: the ordered phases below ~600 K, the order-disordering phase transition temperature range between 800 – 1000 K; and the disordered γ phase above 1200 K. The spatial distribution of the γ"-NbNi$_3$ phase, the α-Cr phase, and the overall morphology at 1200 K, along with the γ-NiFeCr, the γ"-NbNi$_3$ phase, the α-Cr phase, and the overall morphology at 1400 K are plotted in Figure 12. The ordered phases at 600 K mainly consist of the L1$_0$-FeNi (24%), γ'-Ni$_3$Fe (25%), γ'-Ni$_3$Al (4%), γ'-Ni$_3$Ti (4%), Ni$_4$Mo (6%), α-Cr (22%), γ"-NbNi$_3$ (11%), δ-NbNi$_3$ (1%), and δ-MoNi$_3$ (1%). Due to the imposition of the grain boundary constraint, the δ phases have been mostly replaced by the γ" phase away from grain boundaries. In the temperature range between 800 – 1000 K, the ordered L1$_0$ phase is not present, and the amount of the γ'-Ni$_3$Fe phase decreases significantly. The disordered γ phase is mainly a mixture of disordered FeNi, FeNi$_2$, and FeNi$_3$.

When the temperature goes above 1200 K, the amount of α-Cr starts to decrease significantly. The disordered γ phase at 1400 K is mainly made of FeNi$_2$, CrNi$_2$, CrFeNi, and CrMoNi$_3$, with fractions of 12%, 14%, 42%, and 7%, respectively.

5   Conclusions

A comprehensive thermochemical database for multi-component systems involving 26 elements, Ni, Fe, Co, Al, B, C, Cr, Cu, Hf, La, Mn, Mo, N, Nb, O, P, Re, Ru, S, Si, Ta, Ti, V, W, Y, and Zr was developed by a combination of high-throughput 0K DFT calculations and finite temperature lattice dynamics and electron statistics. In particular, the temperature dependencies of chemical potentials (molar Gibbs free energies) of more than 3000 individual phases are computed using the quasiharmonic phonon approach for the lattice contribution and the Mermin statistics for the thermal electronic contribution. In combination with TEM-EDS measurements of local compositional distributions of elements, local phase distributions as well as their temperature dependencies can be obtained from the DFT-derived database, which can be extremely useful information for understanding and controlling the mechanical properties of AM-fabricated superalloys. As an example of application, it is examined to determining the appearance of metastable and stable phases and their spatial distributions in additively manufacturing (AM) fabricated IN718 alloys. In addition to AM-fabricated IN718, this database is directly applicable to many practical alloy systems, e.g. other Ni- and Co-based superalloy systems, such as Hastelloy, Inconel, Waspaloy, Rene alloys, Haynes alloys, Incoloy, MP98T, TMS alloys, and CMSX etc.

**Acknowledgement**: Information within this manuscript is based upon research supported through the Defense Advanced Research Projects Agency's (DARPA's) Open Manufacturing Program.

Any opinions, findings, and conclusions or recommendations expressed in this publication are those of the authors and do not necessarily reflect the views of DARPA (Yi Wang, McNamara, Lia, Martukanitz, and Chen). First-principles calculations were mainly carried out on the resources of NERSC supported by the DOE Office of Science under contract no. DE-AC02-05CH11231 and partially on the resources of Extreme Science and Engineering Discovery Environment (XSEDE) supported by NSF with Grant No. ACI-1053575 (Yi Wang, Shang, Liu, and Chen). This work was partially supported by the Office of Naval Research (ONR) via contract No. N00014-17-1-2567, the U. S. Department of Energy (DOE) via award Nos. DE-FE0031553 and DE-EE0008456, and the National Science Foundation of the United States through CMMI-1825538 (Yi Wang, Shang, and Liu). The authors would like to express our great appreciation to Carpenter Technology Corporation for providing some of the thermodynamic calculations.

Author Contributions: Yi Wang initiated the manuscript, performed first-principles calculation, and participated EDS measurements and result analysis; Frederick Lia prepared the IN718 sample and participated EDS measurement; Ke Wang was in charge of EDS measurements; Kevin McNamara participated EDS measurements and data collections; Yanzhou Ji and Xiaoyu Chong participated in crystal structure data and literature collections; Shun-Li Shang performed initial CALPHAD calculations; Zi-Kui Liu revised the manuscript; Richard P. Martukanitz initiated EDS measurements and revised the manuscript; and Long-Qing Chen was in charge of the overall quality of the work and comprehensively revised the manuscript.

# References

[1] G. Kresse, D. Joubert, From ultrasoft pseudopotentials to the projector augmented-wave method, Phys. Rev. B. 59 (1999) 1758–1775.

[2] G. Kresse, J. Furthmuller, J. Furthmüller, J. Furthmuller, J. Furthmüller, J. Furthmueller, J. Furthmuller, J. Furthmüller, Efficient iterative schemes for ab initio total-energy calculations using a plane-wave basis set, Phys. Rev. B-Condensed Matter. 54 (1996) 11169–11186. doi:10.1103/PhysRevB.54.11169.

[3] W. Kohn, L.J. Sham, SELF-CONSISTENT EQUATIONS INCLUDING EXCHANGE AND CORRELATION EFFECTS, Phys. Rev. 140 (1965) A1133-38.

[4] P. Hohenberg, W. Kohn, Inhomogeneous electron gas, Phys. Rev. 136 (1964) B864.

[5] N.D. Mermin, Thermal properties of the inhomogeneous electron gas, Phys. Rev. 137 (1965) A1441. doi:10.1103/PhysRev.137.A1441.

[6] A. Oradei-Basile, J.F. Radavich, A Current T-T-T Diagram for Wrought Alloy 718, in: E. A Loria (Ed.), The Minerals, Metals & Materials Society, Warrendale, PA, 1991: pp. 325–335. https://www.tms.org/superalloys/10.7449/1991/Superalloys_1991_325_335.pdf (accessed March 2, 2018).

[7] S. Mannan, B. Puckett, Physical Metallurgy of Alloys 718, 925, 725, and 725HS for Service in Aggressive Corrosion Environments, Corros. 2003. (2003) 03126. https://doi.org/.

[8] J.W. Brooks, P.J. Bridges, Metallurgical Stability of Inconel Alloy 718, in: Superalloys 1988 (Sixth Int. Symp., TMS, 1988: pp. 33–42.

doi:10.7449/1988/Superalloys_1988_33_42.

[9]  S.T. Wlodek, R.D. Field, THE EFFECTS OF LONG TIME EXPOSURE ON ALLOY 718, in: E.A. Loria (Ed.), Miner. Met. Soc., 1994: pp. 659–671. http://www.tms.org/Superalloys/10.7449/1994/Superalloys_1994_659_670.pdf (accessed March 2, 2018).

[10] S.P. Ong, W.D. Richards, A. Jain, G. Hautier, M. Kocher, S. Cholia, D. Gunter, V.L. Chevrier, K.A. Persson, G. Ceder, Python Materials Genomics (pymatgen): A robust, open-source python library for materials analysis, Comput. Mater. Sci. 68 (2013) 314–319.

[11] H.T. Stokes, D.M. Hatch, *FINDSYM* : program for identifying the space-group symmetry of a crystal, J. Appl. Crystallogr. 38 (2005) 237–238. doi:10.1107/S0021889804031528.

[12] H.T. Stokes, S. van Orden, B.J. Campbell, *ISOSUBGROUP* : an internet tool for generating isotropy subgroups of crystallographic space groups, J. Appl. Crystallogr. 49 (2016) 1849–1853. doi:10.1107/S160057671601311X.

[13] A. van de Walle, P. Tiwary, M. de Jong, D.L. Olmsted, M. Asta, A. Dick, D. Shin, Y. Wang, L.-Q. Chen, Z.-K. Liu, Efficient stochastic generation of special quasirandom structures, Calphad. 42 (2013) 13–18. doi:10.1016/J.CALPHAD.2013.06.006.

[14] D. Shin, A. van de Walle, Y. Wang, Z.-K. Liu, First-principles study of ternary fcc solution phases from special quasirandom structures, Phys. Rev. B. 76 (2007) 144204. h (accessed June 29, 2017).

[15] A. Zunger, S.-H. Wei, L.G. Ferreira, J.E. Bernard, Special quasirandom structures, Phys. Rev. Lett. 65 (1990) 353–356.

[16] S.-H. Wei, L.G. Ferreira, J.E. Bernard, A. Zunger, Electronic properties of random alloys: Special quasirandom structures, Phys. Rev. B. 42 (1990) 9622–9649. doi:10.1103/PhysRevB.42.9622.

[17] K.C. Hass, L.C. Davis, A. Zunger, Electronic structure of random $Al_{0.5}Ga_{0.5}As$ alloys: Test of the '"special-quasirandom-structures"' description, Phys. Rev. B. 42 (1990) 3757–3760. doi:10.1103/PhysRevB.42.3757.

[18] B. Seiser, R. Drautz, D.G. Pettifor, TCP phase predictions in Ni-based superalloys: Structure maps revisited, Acta Mater. 59 (2011) 749–763. doi:10.1016/J.ACTAMAT.2010.10.013.

[19] R.S. Fishman, S.H. Liu, Magnetic structure and paramagnetic dynamics of chromium and its alloys, Phys. Rev. B. 47 (1993) 11870–11882. doi:10.1103/PhysRevB.47.11870.

[20] V. Sliwko, P. Mohn, K. Schwarz, The electronic and magnetic structures of alpha - and beta -manganese, J. Phys. Condens. Matter. 6 (1994) 6557–6564. doi:10.1088/0953-8984/6/32/017.

[21] J.P. Perdew, K. Burke, M. Ernzerhof, Generalized Gradient Approximation Made Simple, Phys. Rev. Lett. 77 (1996) 3865–3868.

[22] M. Methfessel, A.T. Paxton, High-precision sampling for Brillouin-zone integration in metals, Phys. Rev. B. 40 (1989) 3616–3621. doi:10.1103/PhysRevB.40.3616.

[23] P.E. Blöchl, O. Jepsen, O.K. Andersen, Improved tetrahedron method for Brillouin-zone integrations, Phys. Rev. B. 49 (1994) 16223–16233. doi:10.1103/PhysRevB.49.16223.

[24] Y. Wang, L.-Q. Chen, Z.-K. Liu, YPHON: A package for calculating phonons of polar materials, Comput. Phys. Commun. 185 (2014) 2950–2968.

[25] Y. Wang, J.J. Wang, W.Y. Wang, Z.G. Mei, S.L. Shang, L.Q. Chen, Z.K. Liu, A Mixed-space Approach to First-principles Calculations of Phonon Frequencies for Polar Materials, J. Physics-Condensed Matter. 22 (2010) 202201.

[26] Z.-K. Liu, Y. Wang, Computational thermodynamics of materials, Cambridge University Press, Cambridge, UK, 2016.

[27] Y. Wang, S.-L. Shang, H. Fang, Z.-K. Liu, L.-Q. Chen, First-principles calculations of lattice dynamics and thermal properties of polar solids, Npj Comput. Mater. 2 (2016) 16006. doi:10.1038/npjcompumats.2016.6.

[28] L.Q. Chen, Chemical potential and Gibbs free energy, MRS Bull. 44 (2019) 520–523. doi:10.1557/mrs.2019.162.

[29] Y. Wang, Z.-K. Liu, L.-Q. Chen, Thermodynamic properties of Al, Ni, NiAl, and Ni3Al from first-principles calculations, Acta Mater. 52 (2004) 2665–2671. doi:10.1016/J.ACTAMAT.2004.02.014.

[30] Y. Wang, J.J. Wang, H. Zhang, V.R. Manga, S.L. Shang, L.-Q. Chen, Z.-K. Liu, A first-principles approach to finite temperature elastic constants, J. Phys. Condens. Matter. 22 (2010) 225404. doi:10.1088/0953-8984/22/22/225404.

[31] Y. Wang, Classical mean-field approach for thermodynamics: Ab initio thermophysical properties of cerium, Phys. Rev. B. 61 (2000) R11863.

[32] Y. Wang, L. Li, Mean-field potential approach to thermodynamic properties of metal: Al as a prototype, Phys. Rev. B. 62 (2000) 196–202.

[33] R.P.P. Stoffel, C. Wessel, M.-W. Lumey, R. Dronskowski, Ab Initio Thermochemistry of Solid-State Materials, Angew. Chemie Int. Ed. 49 (2010) 5242–5266.

doi:10.1002/anie.200906780.

[34] J.M. Skelton, S.C. Parker, A. Togo, I. Tanaka, A. Walsh, Thermal physics of the lead chalcogenides PbS, PbSe, and PbTe from first principles, Phys. Rev. B. 89 (2014) 205203. doi:10.1103/PhysRevB.89.205203.

[35] Y. Wang, R. Ahuja, B. Johansson, Mean-field potential approach to the quasiharmonic theory of solids, Int. J. Quantum Chem. 96 (2004) 501–506. doi:10.1002/qua.10769.

[36] V.L. Moruzzi, J.F. Janak, K. Schwarz, Calculated thermal properties of metals, Phys. Rev. B. 37 (1988) 790–799.

[37] S.-L. Shang, Y. Wang, D. Kim, Z.-K. Liu, First-principles thermodynamics from phonon and Debye model: Application to Ni and Ni3Al, Comput. Mater. Sci. 47 (2010) 1040–1048. doi:10.1016/J.COMMATSCI.2009.12.006.

[38] A.K. McMahan, M. Ross, High-temperature electron-band calculations, Phys. Rev. B. 15 (1977) 718.

[39] X. Wang, X. Gong, K. Chou, Review on powder-bed laser additive manufacturing of Inconel 718 parts, Proc. Inst. Mech. Eng. Part B J. Eng. Manuf. 231 (2016) 1890–1903. doi:10.1177/0954405415619883.

[40] R.C. Reed, The Superalloys: Fundamentals and Applications, Cambridge University Press, Cambridge, 2006.

[41] T.M. Pollock, S. Tin, Nickel-Based Superalloys for Advanced Turbine Engines: Chemistry, Microstructure and Properties, J. Propuls. Power. 22 (2006) 361–374. doi:10.2514/1.18239.

[42] M.J. Donachie, S.J. Donachie, Superalloys : a technical guide, ASM International, 2002.

https://www.asminternational.org/technical-books/-/journal_content/56/10192/06128G/PUBLICATION (accessed February 12, 2019).

[43] M. Jambor, O. Bokůvka, F. Nový, L. Trško, J. Belan, Phase transformations in nickel base superalloy INCONEL 718 during cyclic loading at high temperature, Prod. Eng. Arch. 15 (2017) 15–18. doi:10.30657/pea.2017.15.04.

[44] M. Sundararaman, P. Mukhopadhyay, S. Banerjee, CARBIDE PRECIPITATION IN NICKEL BASE SUPERALLOYS 718 AND 625 AND THEIR EFFECT ON MECHANICAL PROPERTIES, in: E.A. Loria (Ed.), Superalloys 718,625,706 Var. Deriv., The Minerals, Metals &Materials Society, 1997: pp. 367–378. https://www.tms.org/Superalloys/10.7449/1997/Superalloys_1997_367_378.pdf (accessed July 17, 2019).

[45] P. Promoppatum, S.-C. Yao, P. Chris Pistorius, A.D. Rollett, P.J. Coutts, F. Lia, R. Martukanitz, Numerical modeling and experimental validation of thermal history and microstructure for additive manufacturing of an Inconel 718 product, Prog. Addit. Manuf. 3 (2018) 15–32. doi:10.1007/s40964-018-0039-1.

[46] A. Keshavarzkermani, M. Sadowski, L. Ladani, Direct metal laser melting of Inconel 718: Process impact on grain formation and orientation, J. Alloys Compd. 736 (2018) 297–305. doi:10.1016/j.jallcom.2017.11.130.

[47] H. Helmer, A. Bauereiß, R.F. Singer, C. Körner, Grain structure evolution in Inconel 718 during selective electron beam melting, Mater. Sci. Eng. A. 668 (2016) 180–187. doi:10.1016/j.msea.2016.05.046.

[48] N. Raghavan, S. Simunovic, R. Dehoff, A. Plotkowski, J. Turner, M. Kirka, S. Babu,

https://www.asminternational.org/technical-books/-/journal_content/56/10192/06128G/PUBLICATION (accessed February 12, 2019).

[43] M. Jambor, O. Bokůvka, F. Nový, L. Trško, J. Belan, Phase transformations in nickel base superalloy INCONEL 718 during cyclic loading at high temperature, Prod. Eng. Arch. 15 (2017) 15–18. doi:10.30657/pea.2017.15.04.

[44] M. Sundararaman, P. Mukhopadhyay, S. Banerjee, CARBIDE PRECIPITATION IN NICKEL BASE SUPERALLOYS 718 AND 625 AND THEIR EFFECT ON MECHANICAL PROPERTIES, in: E.A. Loria (Ed.), Superalloys 718,625,706 Var. Deriv., The Minerals, Metals &Materials Society, 1997: pp. 367–378. https://www.tms.org/Superalloys/10.7449/1997/Superalloys_1997_367_378.pdf (accessed July 17, 2019).

[45] P. Promoppatum, S.-C. Yao, P. Chris Pistorius, A.D. Rollett, P.J. Coutts, F. Lia, R. Martukanitz, Numerical modeling and experimental validation of thermal history and microstructure for additive manufacturing of an Inconel 718 product, Prog. Addit. Manuf. 3 (2018) 15–32. doi:10.1007/s40964-018-0039-1.

[46] A. Keshavarzkermani, M. Sadowski, L. Ladani, Direct metal laser melting of Inconel 718: Process impact on grain formation and orientation, J. Alloys Compd. 736 (2018) 297–305. doi:10.1016/j.jallcom.2017.11.130.

[47] H. Helmer, A. Bauereiß, R.F. Singer, C. Körner, Grain structure evolution in Inconel 718 during selective electron beam melting, Mater. Sci. Eng. A. 668 (2016) 180–187. doi:10.1016/j.msea.2016.05.046.

[48] N. Raghavan, S. Simunovic, R. Dehoff, A. Plotkowski, J. Turner, M. Kirka, S. Babu,

Localized melt-scan strategy for site specific control of grain size and primary dendrite arm spacing in electron beam additive manufacturing, Acta Mater. 140 (2017) 375–387. doi:10.1016/j.actamat.2017.08.038.

[49] A. Mostafa, I. Picazo Rubio, V. Brailovski, M. Jahazi, M. Medraj, Structure, Texture and Phases in 3D Printed IN718 Alloy Subjected to Homogenization and HIP Treatments, Metals (Basel). 7 (2017) 196. doi:10.3390/met7060196.

[50] A. Thomas, M. El-Wahabi, J.M. Cabrera, J.M. Prado, High temperature deformation of Inconel 718, J. Mater. Process. Technol. 177 (2006) 469–472. doi:10.1016/J.JMATPROTEC.2006.04.072.

[51] J.A.D. Connolly, K. Petrini, An automated strategy for calculation of phase diagram sections and retrieval of rock properties as a function of physical conditions, J. Metamorph. Geol. 20 (2002) 697–708.

[52] G.B. Dantzig, A. Orden, P. Wolfe, A.P. Dempster, S. Schuster, H.L. Royden, R.P. Dilworth, E. Hewitt, A. Horn, H. Busemann, P.R. Halmos, R.D. James, G. Polya, H. Federer, H. Hopf, B. Jessen, J.J. Stoker, M. Hall, A. Horn, P. Levy, K. Yosida, L.M. Blumenthal, The generalized simplex method for minimizing a linear form under linear inequality restraints, Pacific J. Math. 5 (1955) 183–195.

[53] E.L. Stevens, J. Toman, A.C. To, M. Chmielus, Variation of hardness, microstructure, and Laves phase distribution in direct laser deposited alloy 718 cuboids, Mater. Des. 119 (2017) 188–198. doi:10.1016/j.matdes.2017.01.031.

[54] S. Sui, J. Chen, E. Fan, H. Yang, X. Lin, W. Huang, The influence of Laves phases on the high-cycle fatigue behavior of laser additive manufactured Inconel 718, Mater. Sci. Eng.

A. 695 (2017) 6–13. doi:10.1016/j.msea.2017.03.098.

[55] S. Sui, J. Chen, X. Ming, S. Zhang, X. Lin, W. Huang, The failure mechanism of 50% laser additive manufactured Inconel 718 and the deformation behavior of Laves phases during a tensile process, Int. J. Adv. Manuf. Technol. 91 (2017) 2733–2740. doi:10.1007/s00170-016-9901-9.

[56] H. Xiao, S. Li, X. Han, J. Mazumder, L. Song, Laves phase control of Inconel 718 alloy using quasi-continuous-wave laser additive manufacturing, Mater. Des. 122 (2017) 330–339. doi:10.1016/j.matdes.2017.03.004.

[57] A. Yadollahi, N. Shamsaei, Additive manufacturing of fatigue resistant materials: Challenges and opportunities, Int. J. Fatigue. 98 (2017) 14–31. doi:10.1016/j.ijfatigue.2017.01.001.

[58] Y. Idell, L.E. Levine, A.J. Allen, F. Zhang, C.E. Campbell, G.B. Olson, J. Gong, D.R. Snyder, H.Z. Deutchman, Unexpected δ-Phase Formation in Additive-Manufactured Ni-Based Superalloy, Jom. 68 (2016) 950–959. doi:10.1007/s11837-015-1772-2.

[59] Y.-L. Kuo, S. Horikawa, K. Kakehi, The effect of interdendritic δ phase on the mechanical properties of Alloy 718 built up by additive manufacturing, Mater. Des. 116 (2017) 411–418. doi:10.1016/j.matdes.2016.12.026.

[60] J.-O. Andersson, T. Helander, L. Höglund, P. Shi, B. Sundman, Thermo-Calc & DICTRA, computational tools for materials science, Calphad. 26 (2002) 273–312.

[61] J. Bratberg, H. Mao, L. Kjellqvist, A. Engström, P. Mason, Q. Chen, The Development and Validation of a New Thermodynamic Database for Ni-Based Alloys, in: Superalloys 2012, John Wiley & Sons, Inc., Hoboken, NJ, USA, 2012: pp. 803–812.

doi:10.1002/9781118516430.ch89.

[62] H. Hosaeus, A. Seifter, E. Kaschnitz, G. Pottlacher, Thermophysical properties of solid and liquid Inconel 718 alloy, High Temp. High Press. 33 (2001) 405–410.

[63] K.C. Mills, Recommended values of thermophysical properties for selected commercial alloys, Woodhead, Cambridge, England, 2002.

[64] D. Basak, R.A. Overfelt, D. Wang, Measurement of Specific Heat Capacity and Electrical Resistivity of Industrial Alloys Using Pulse Heating Techniques, Int. J. Thermophys. 24 (2003) 1721–1733. doi:10.1023/B:IJOT.0000004101.88449.86.

[65] C.R. Brooks, M. Cash, A. Garcia, The heat capacity of inconel 718 from 313 to 1053 K, J. Nucl. Mater. 78 (1978) 419–421. doi:10.1016/0022-3115(78)90463-4.

[66] S.H. Lee, S.W. Kim, K.H. Kang, Effect of Heat Treatment on the Specific Heat Capacity of Nickel-Based Alloys, Int. J. Thermophys. 27 (2006) 282–292. doi:10.1007/s10765-006-0029-2.

[67] C.-W. Yang, D.B. Williams, J.I. Goldstein, A revision of the Fe-Ni phase diagram at low temperatures (<400 °C), J. Phase Equilibria. 17 (1996) 522–531. doi:10.1007/BF02665999.

[68] J. Yang, J.I. Goldstein, The formation of the Widmanstätten structure in meteorites, Meteorit. Planet. Sci. 40 (2005) 239–253. doi:10.1111/j.1945-5100.2005.tb00378.x.

[69] F.R. Caliari, N.M. Guimarães, D.A.P. Reis, A.A. Couto, C. de Moura Neto, K.C.G. Candioto, Study of the Secondary Phases in Inconel 718 Aged Superalloy Using Thermodynamics Modeling, Key Eng. Mater. 553 (2013) 23–28. doi:10.4028/www.scientific.net/KEM.553.23.

[70] M. Sundararaman, P. Mukhopadhyay, S. Banerjee, Some aspects of the precipitation of metastable intermetallic phases in INCONEL 718, Metall. Trans. A. 23 (1992) 2015–2028. doi:10.1007/BF02647549.

[71] J. Kim, S. Kim, J.-Y. Suh, Y.J. Kim, Y.K. Kim, H. Choi-Yim, Properties of a rare earth free L10-FeNi hard magnet developed through annealing of FeNiPC amorphous ribbons, Curr. Appl. Phys. 19 (2019) 599–605. doi:10.1016/J.CAP.2019.03.001.

[72] E. Poirier, F.E. Pinkerton, R. Kubic, R.K. Mishra, N. Bordeaux, A. Mubarok, L.H. Lewis, J.I. Goldstein, R. Skomski, K. Barmak, Intrinsic magnetic properties of $L1_0$ FeNi obtained from meteorite NWA 6259, J. Appl. Phys. 117 (2015) 17E318. doi:10.1063/1.4916190.

[73] W.F. Hunt, Tetrataenite—ordered FeNi, a new mineral in meteorites, Am. Mineral. 65 (1980) 624–630. https://pubs.geoscienceworld.org/msa/ammin/article/65/7-8/624/41156/tetrataenite-ordered-feni-a-new-mineral-in.

[74] N. Bordeaux, A.M. Montes-Arango, J. Liu, K. Barmak, L.H. Lewis, Thermodynamic and kinetic parameters of the chemical order–disorder transformation in L10 FeNi (tetrataenite), Acta Mater. 103 (2016) 608–615. doi:10.1016/J.ACTAMAT.2015.10.042.

[75] Y. Mishin, M.J. Mehl, D.A. Papaconstantopoulos, Phase stability in the Fe–Ni system: Investigation by first-principles calculations and atomistic simulations, Acta Mater. 53 (2005) 4029–4041. doi:10.1016/J.ACTAMAT.2005.05.001.

[76] G. Cacciamani, J. De Keyzer, R. Ferro, U.E. Klotz, P. Wollants, Critical evaluation of the Fe–Ni, Fe–Ti and Fe–Ni–Ti alloy systems, Intermetallics. 14 (2006) 1312–1325. doi:10.1016/J.INTERMET.2005.11.028.

[77]   L.J. Swartzendruber, V.P. Itkin, C.B. Alcock, The Fe-Ni (iron-nickel) system, J. Phase Equilibria. 12 (1991) 288–312. doi:10.1007/BF02649918.

[78]   G. Cacciamani, J. De Keyzer, R. Ferro, U.E. Klotz, P. Wollants, Critical evaluation of the Fe–Ni, Fe–Ti and Fe–Ni–Ti alloy systems, Intermetallics. 14 (2006) 1312–1325. doi:10.1016/J.INTERMET.2005.11.028.

[79]   J. Liu, L.J. Riddiford, C. Floristean, F. Goncalves-Neto, M. Rezaeeyazdi, L.H. Lewis, K. Barmak, Kinetics of order-disorder transformation of L12 FeNi3 in the Fe-Ni system, J. Alloys Compd. 689 (2016) 593–598. doi:10.1016/J.JALLCOM.2016.08.036.

[80]   J.S. Wróbel, D. Nguyen-Manh, M.Y. Lavrentiev, M. Muzyk, S.L. Dudarev, Phase stability of ternary fcc and bcc Fe-Cr-Ni alloys, Phys. Rev. B. 91 (2015) 024108. doi:10.1103/PhysRevB.91.024108.

[81]   G. Cacciamani, A. Dinsdale, M. Palumbo, A. Pasturel, The Fe–Ni system: Thermodynamic modelling assisted by atomistic calculations, Intermetallics. 18 (2010) 1148–1162. doi:10.1016/J.INTERMET.2010.02.026.

[82]   N. Bordeaux, A.M. Montes-Arango, J. Liu, K. Barmak, L.H. Lewis, Thermodynamic and kinetic parameters of the chemical order–disorder transformation in L10 FeNi (tetrataenite), Acta Mater. 103 (2016) 608–615. doi:10.1016/J.ACTAMAT.2015.10.042.

[83]   S.H. Zhou, Y. Wang, C. Jiang, J.Z. Zhu, L.-Q. Chen, Z.-K. Liu, First-principles calculations and thermodynamic modeling of the Ni–Mo system, Mater. Sci. Eng. A. 397 (2005) 288–296. doi:10.1016/J.MSEA.2005.02.037.

[84]   H. Chen, Y. Du, Refinement of the thermodynamic modeling of the Nb–Ni system, Calphad. 30 (2006) 308–315. doi:10.1016/J.CALPHAD.2006.02.005.

[85]  Z. Bi, J. Dong, M. Zhang, L. Zheng, X. Xie, Mechanism of α-Cr precipitation and crystallographic relationships between α-Cr and δ phases in Inconel 718 alloy after long-time thermal exposure, Int. J. Miner. Metall. Mater. 17 (2010) 312–317. doi:10.1007/s12613-010-0310-z.

**Figure captions**

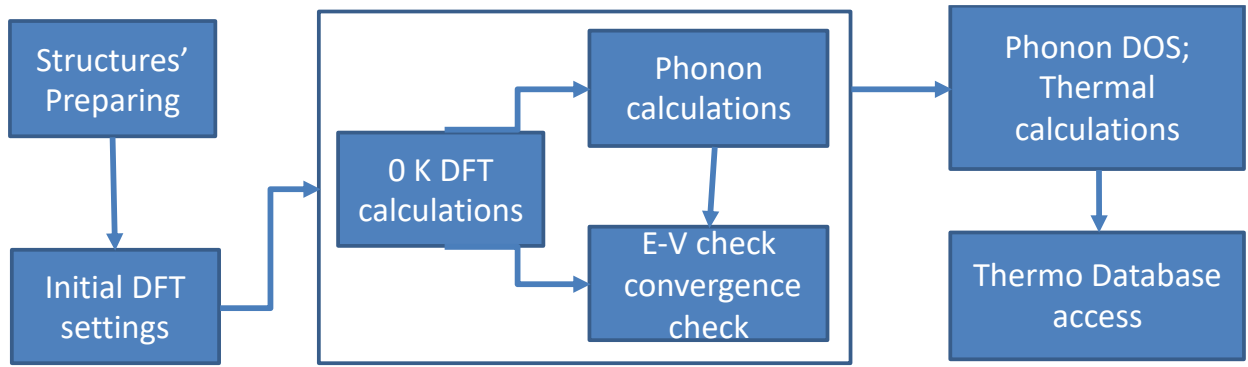

Figure 1. Flowchart for high-throughput quasiharmonic calculations.

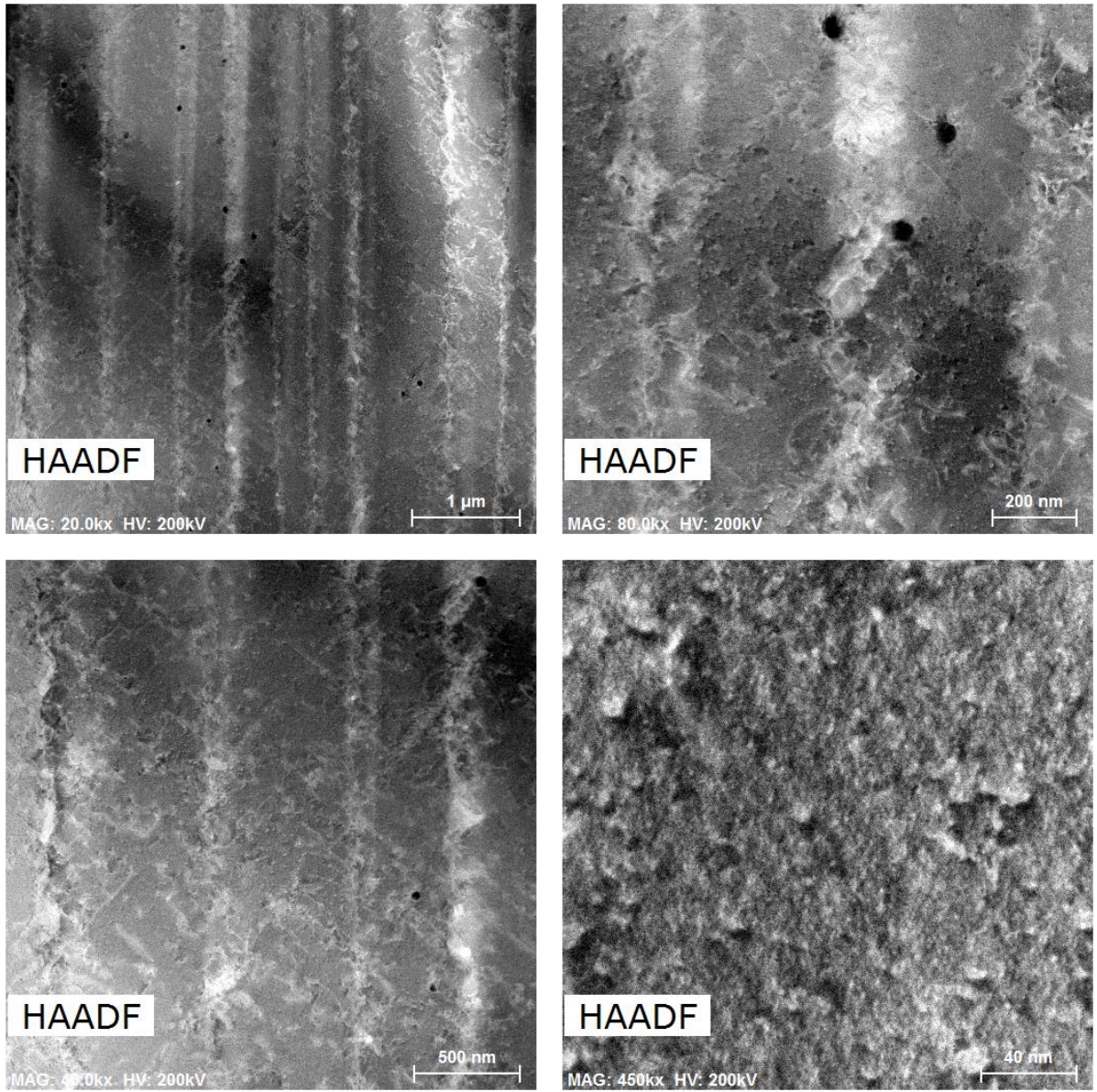

Figure 2. HAADF images of at the scales of 1 μm, 500 nm, 200 nm, and 40 nm.

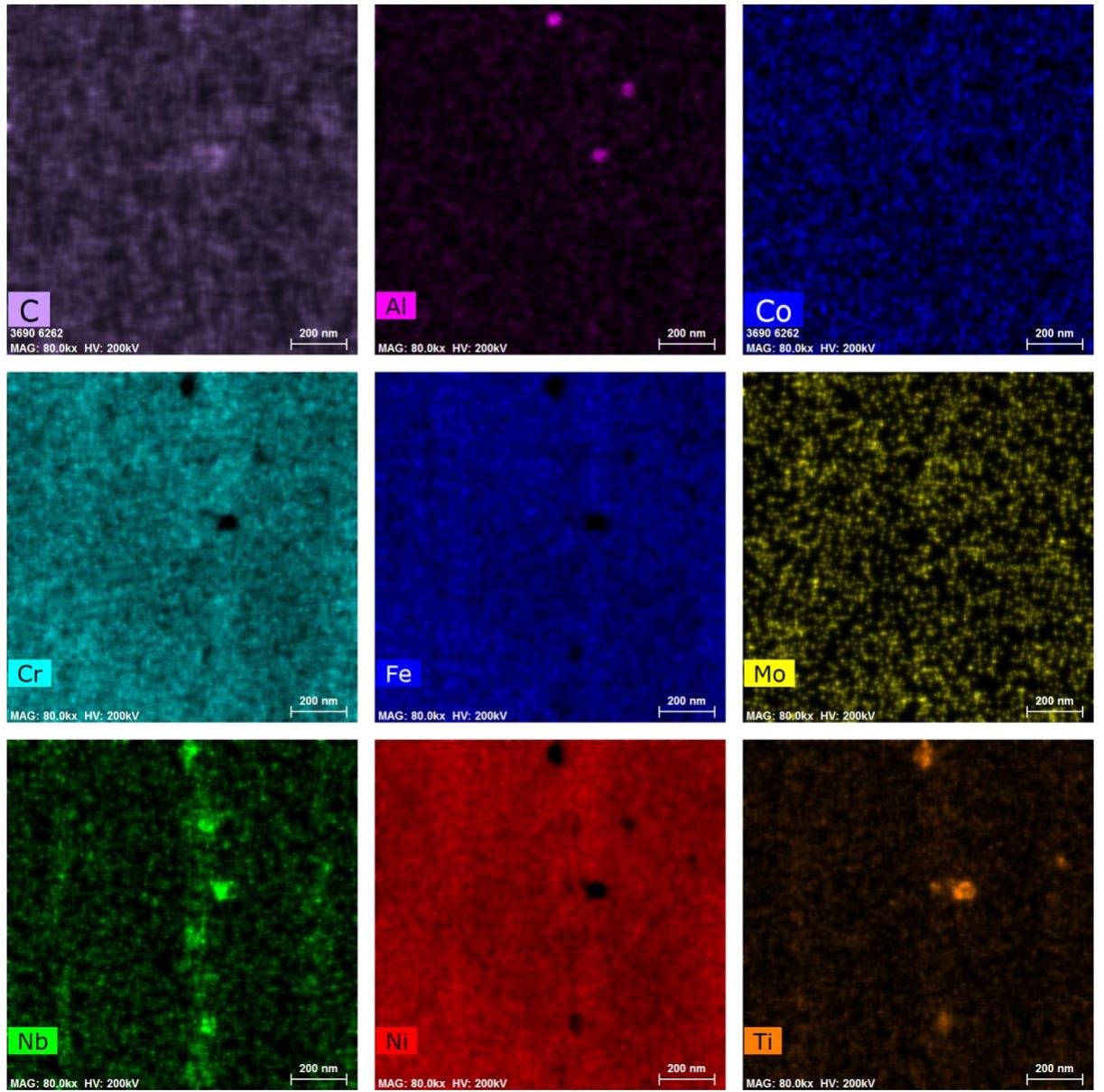

Figure 3. EDS map for the measurement at 200 nm scale.

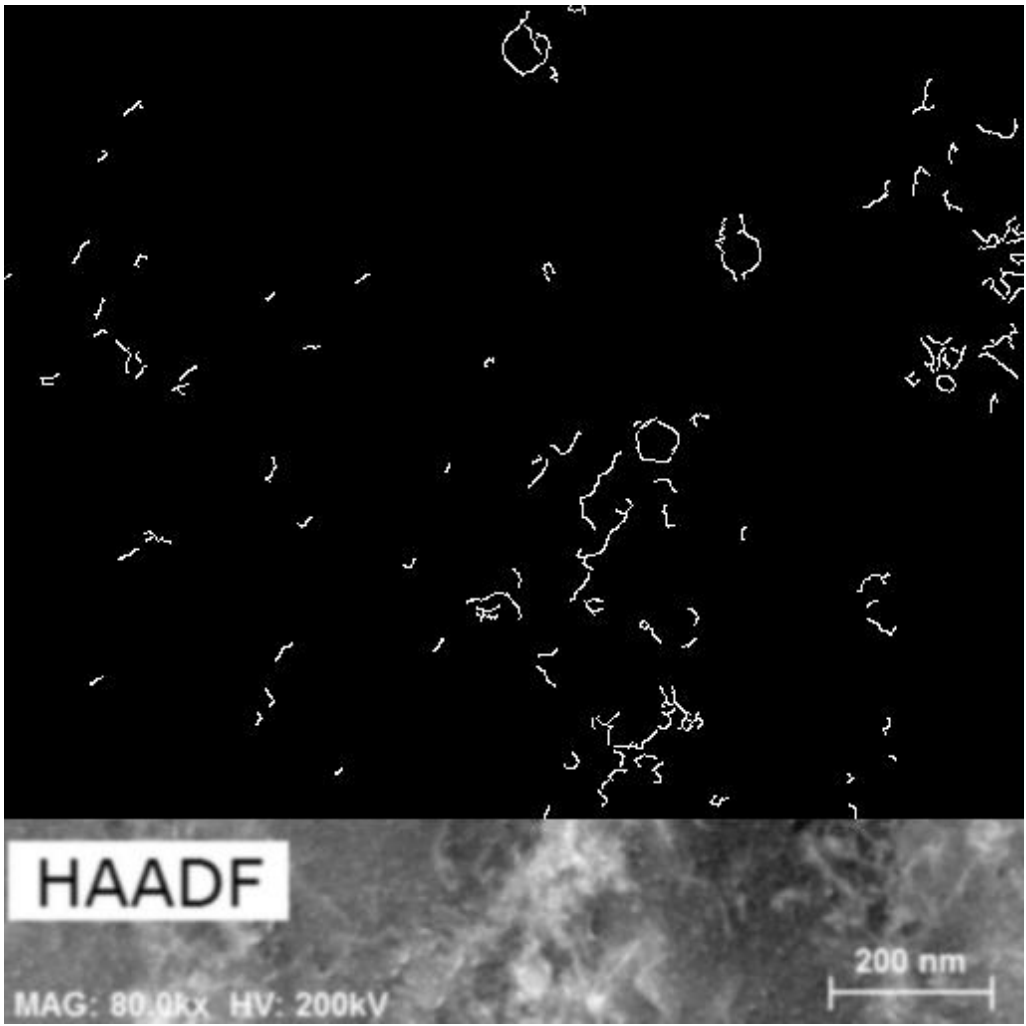

Figure 4. Assumed grain boundaries for the measurement at 200 nm scale.

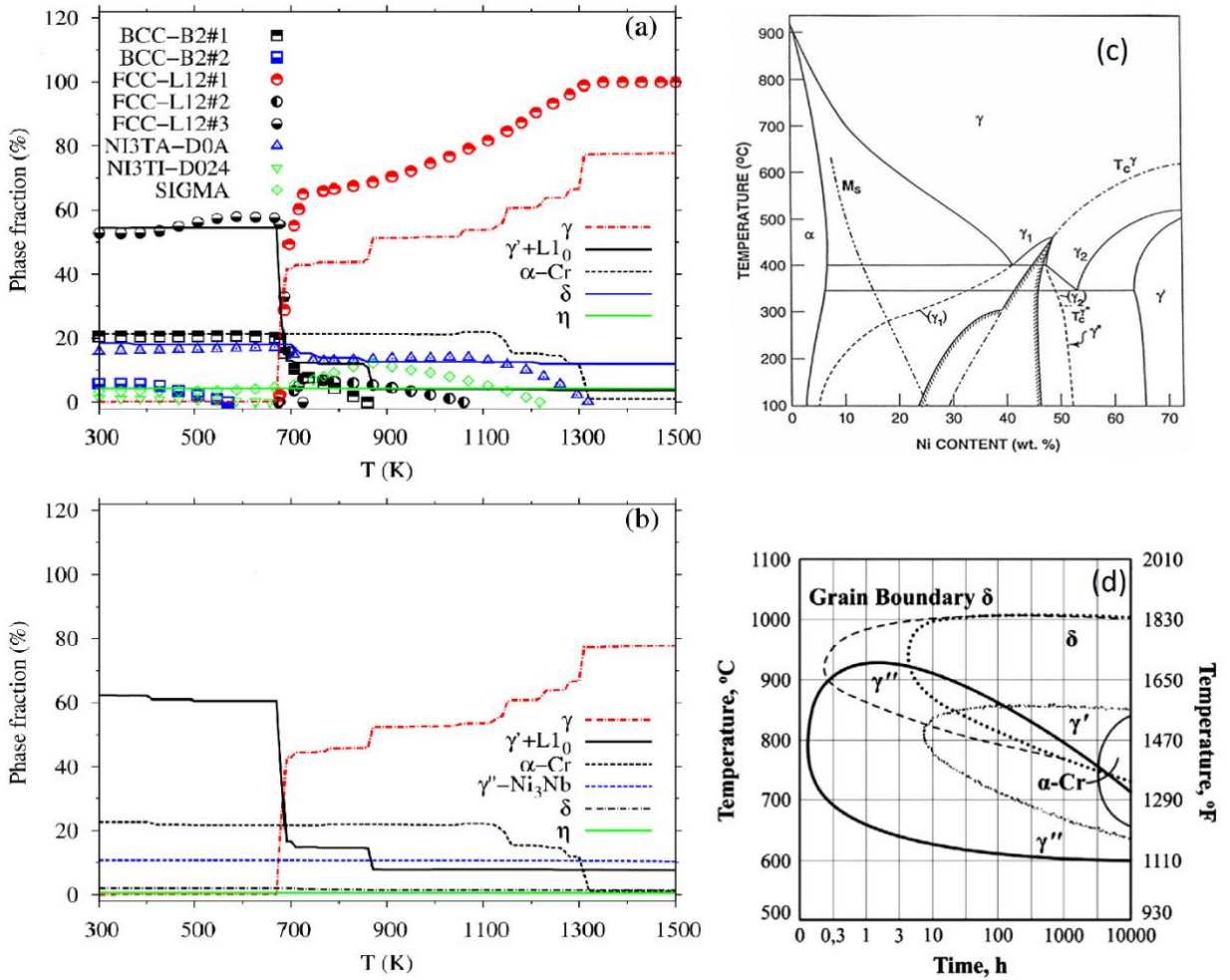

Figure 5. (a) phase fractions for IN718 calculated by DFT (lines) without considering grain boundary constraint compared those calculated based on TCNI9 (symbols). (b) phase fractions for IN718 calculated by DFT (lines) while considering grain boundary constraint. (c) Fe-Ni binary phase diagram (Yang et al. [67,68]). (d) T-T-T diagram for IN718 (Oradei-Basile and Radavich [6]).

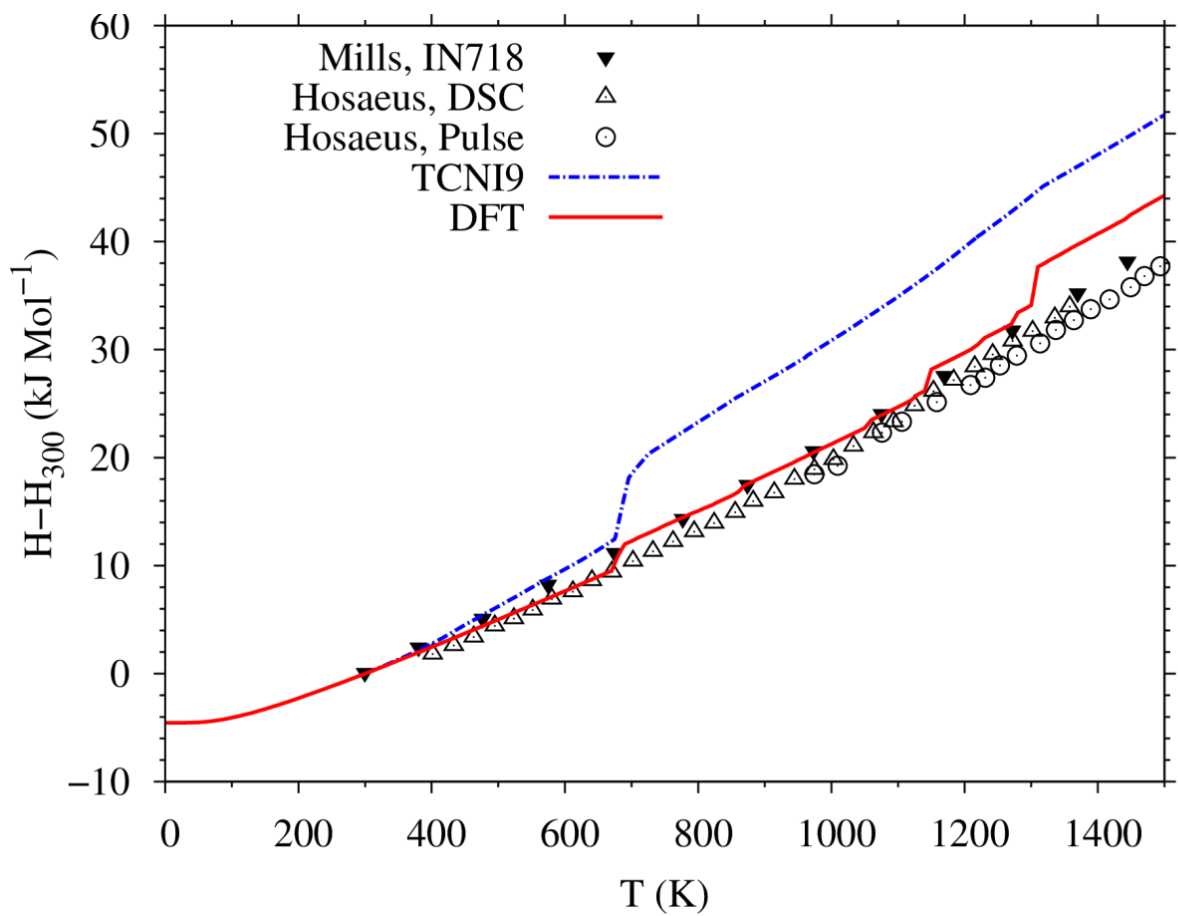

Figure 6. Enthalpies of IN718 as a function of temperature. The solid line represents the DFT calculation and the dot-dashed line represents the calculations based on TCNI9. The solid triangles represent the recommended data by Mills [63] and the open triangles and circles represent the experimental data of Hosaeus et al. [62].

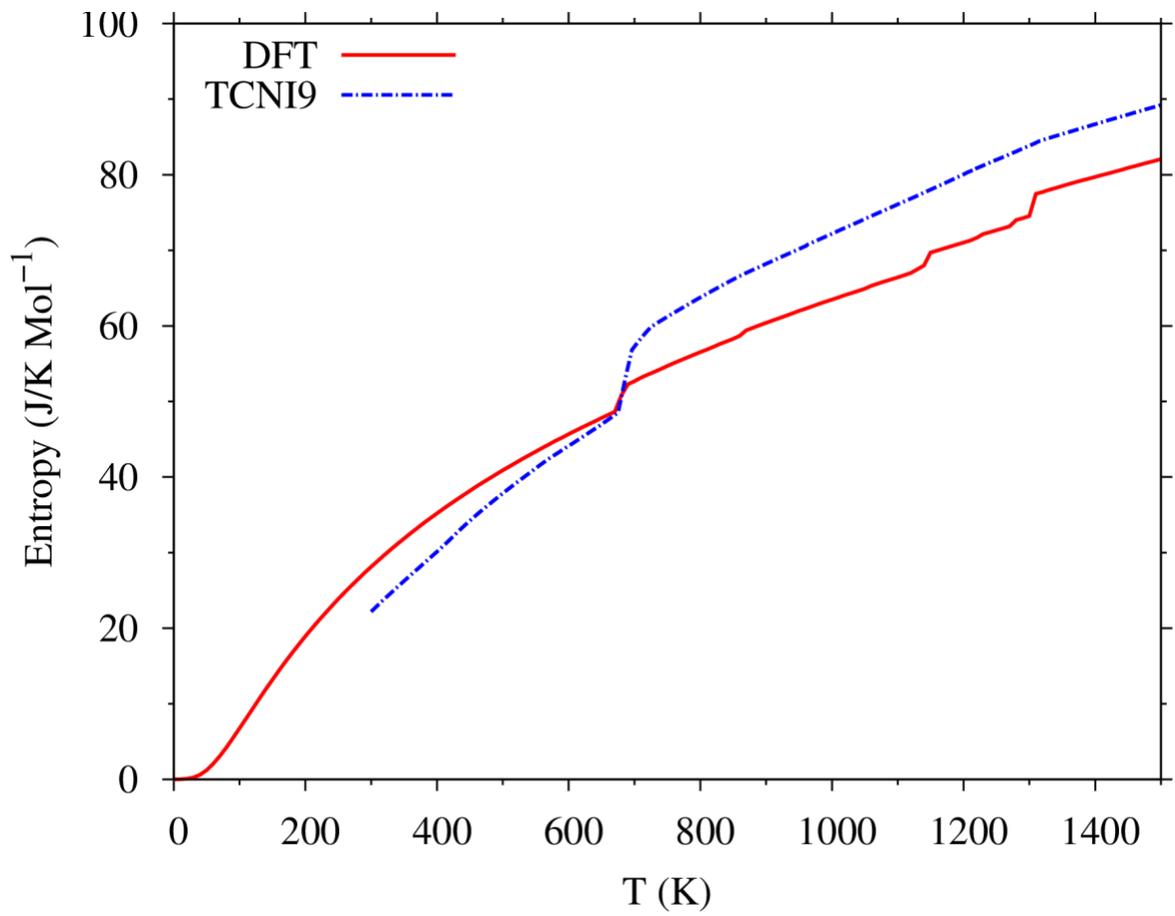

Figure 7. Entropy of IN718 as a function of temperature. The solid line represents the DFT calculation and the dot-dashed line represents the calculations based on TCNI9.

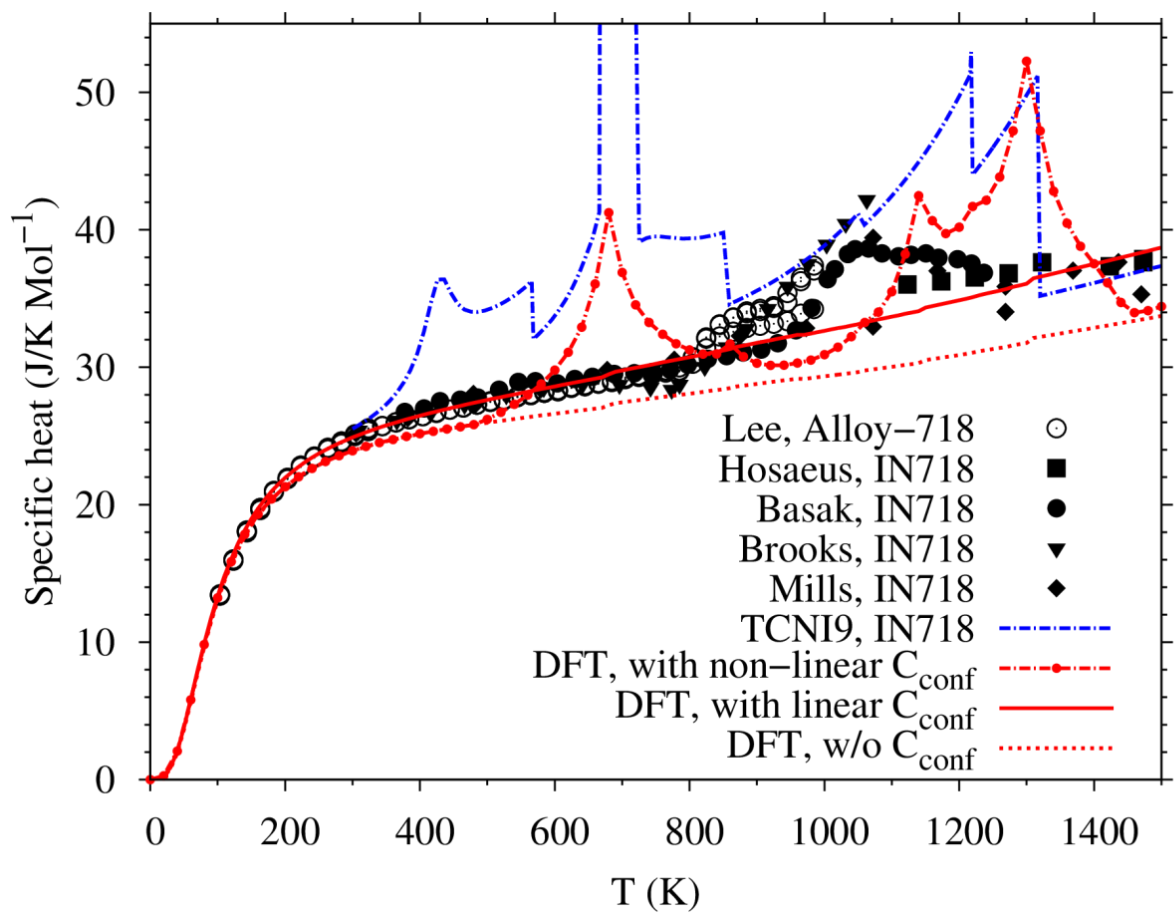

Figure 8. Heat capacities of IN718 as a function of temperature. Open circle: the measured data for Alloy-718 by Lee et al. [66], solid squares: measured data for IN718 from Hosaeus et al. [62], solid circles: measured data for IN718 from Basak et al. [64], solid triangles: measured data for IN718 from Brooks et al. [65], solid diamonds: recommended data for IN718 by Mills [63], dot-dashed line: the calculations based on TCNI9, dot-dashed line with small solid circles: DFT calculation with average scheme#2 (see main text), solid line: DFT calculation with average scheme#1 (see main text), and dotted line: DFT calculation without considering configurational contribution.

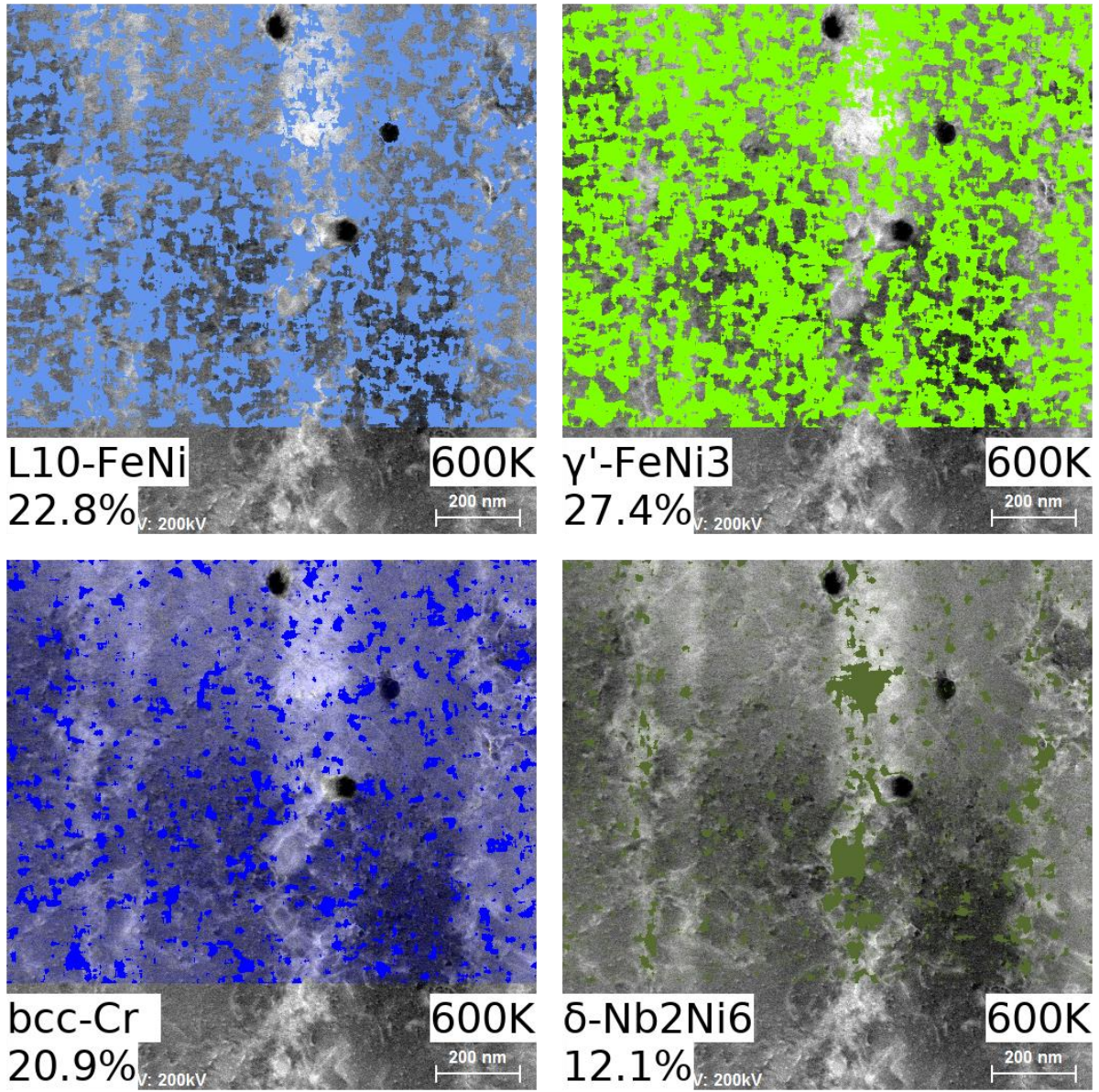

Figure 9. Calculated phase distributions of L1$_0$, γ', bcc-Cr (α-Cr), and δ phases at 600 K without considering grain boundary constraint.

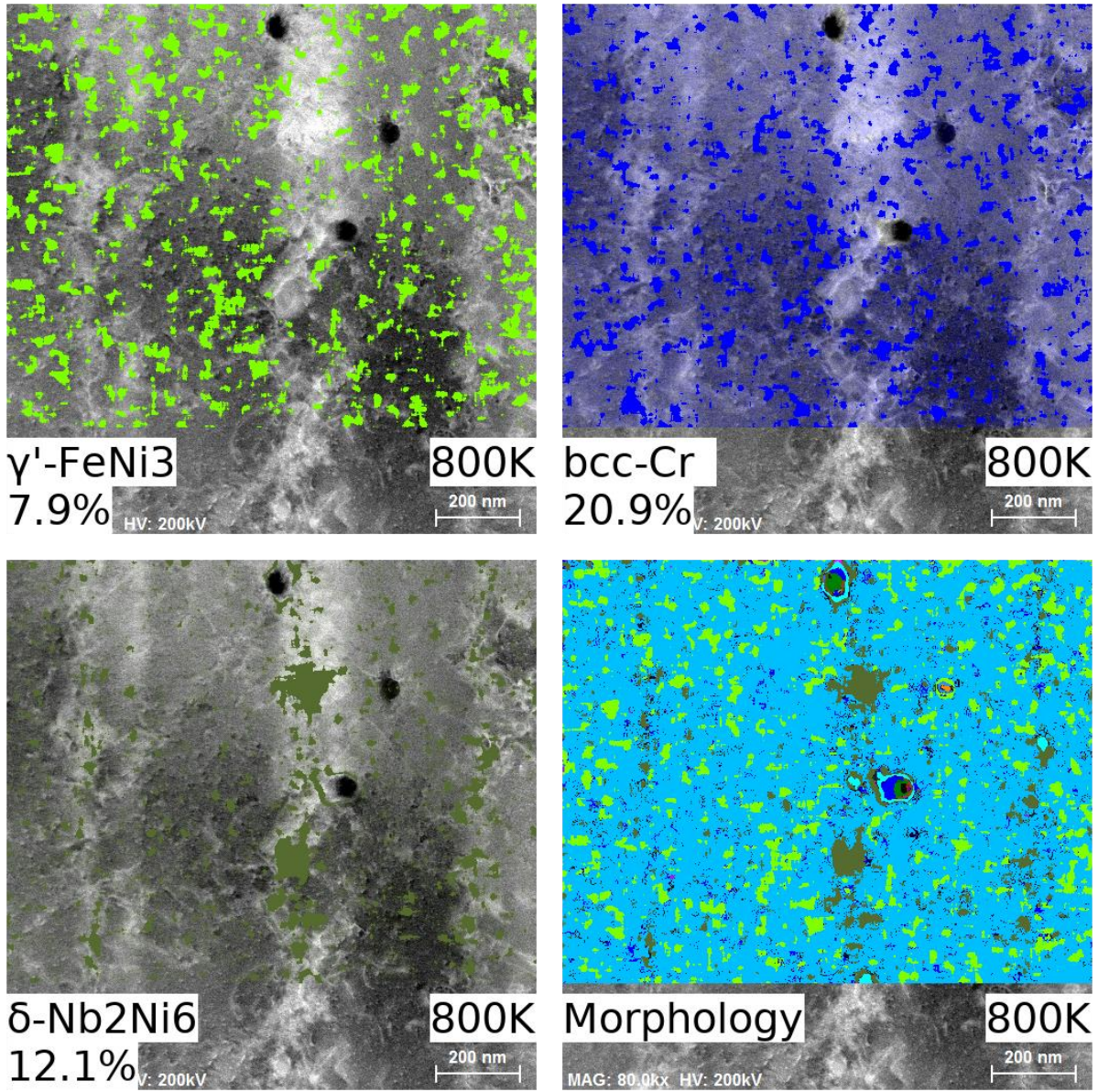

Figure 10. Calculated phase distributions of γ', bcc-Cr (α-Cr), and δ phases, and overall morphology at 800 K without considering grain boundary constraint.

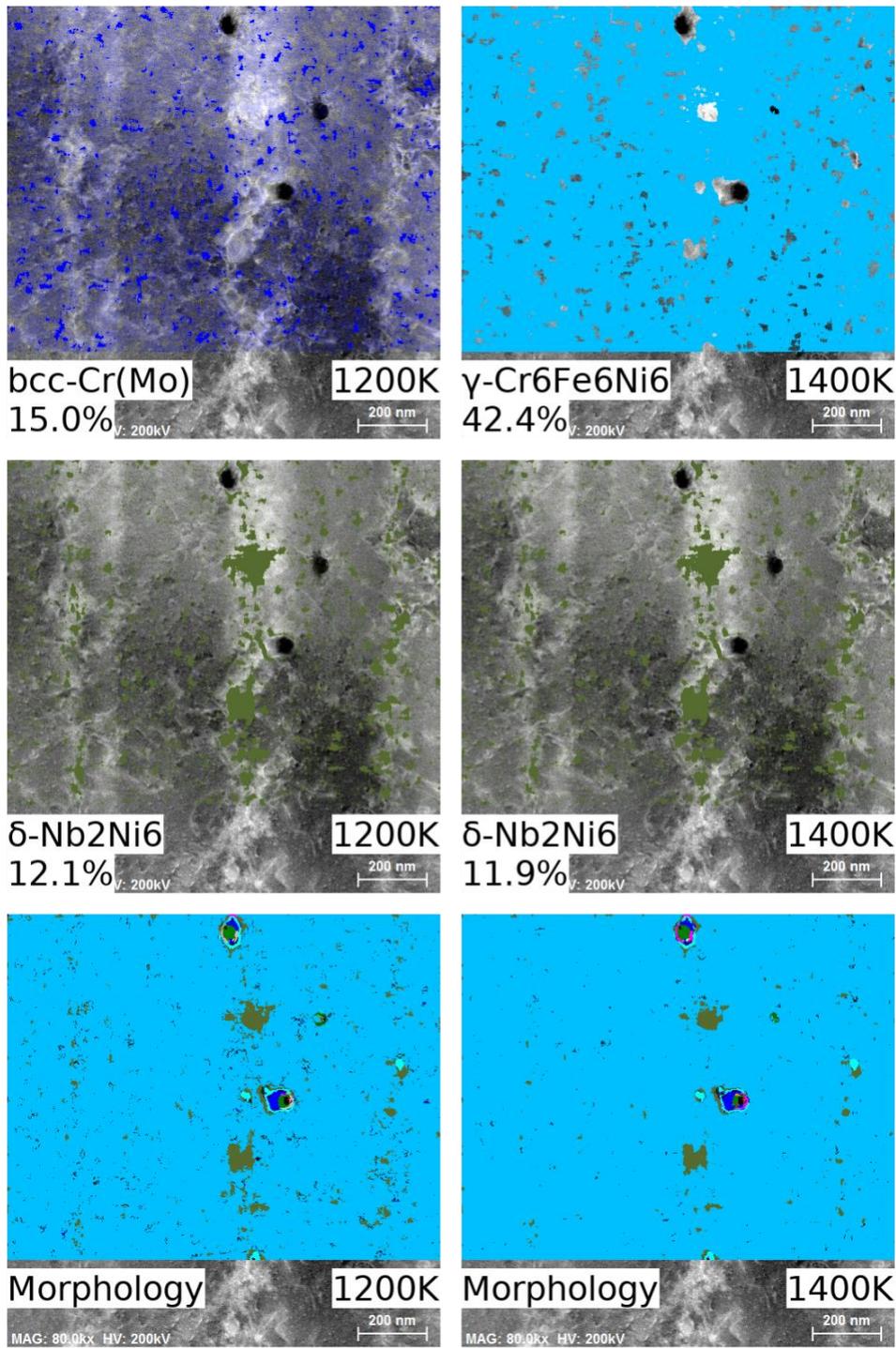

Figure 11. Calculated phase distribution at 1200 K and 1400 K without considering grain boundary constraint.

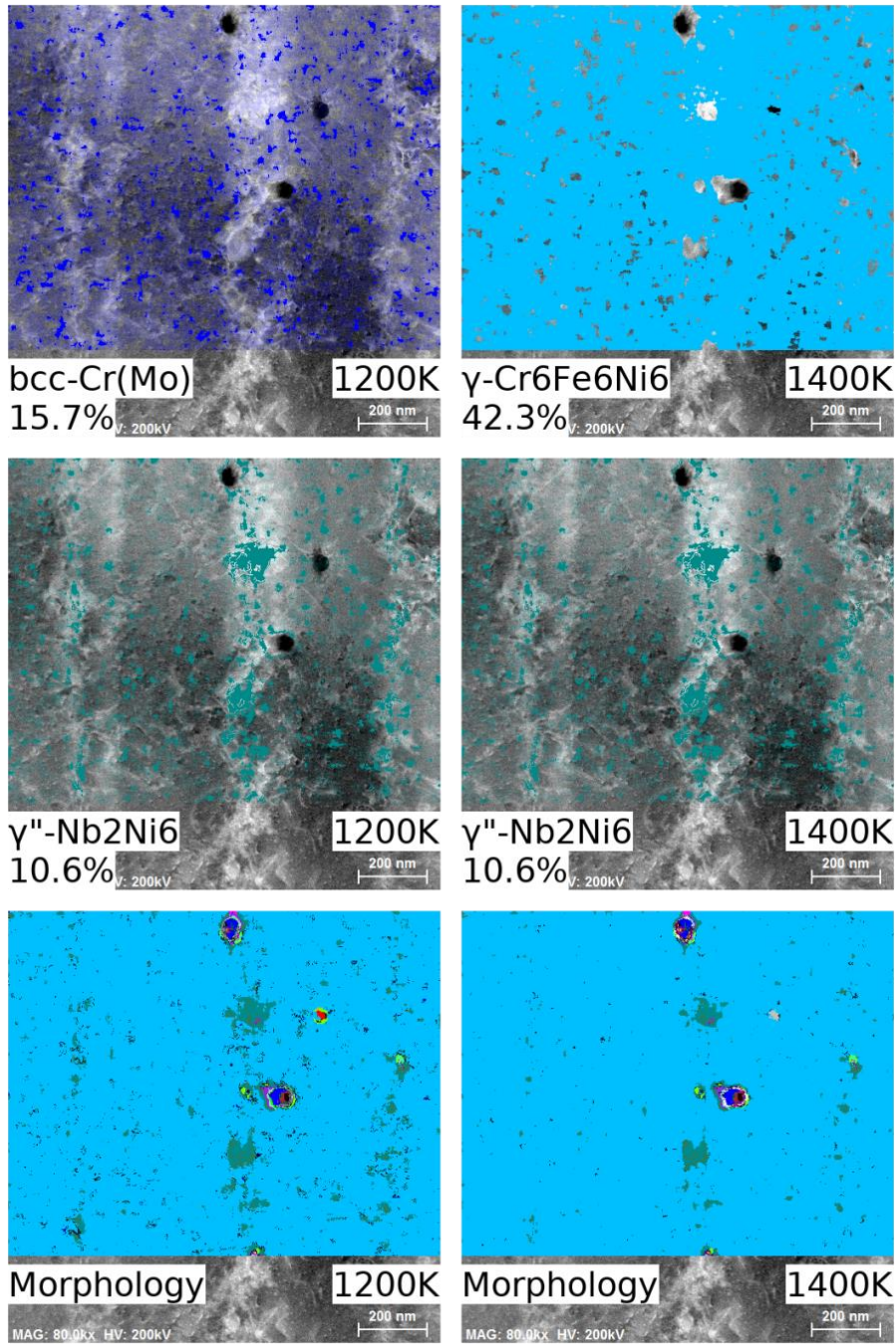

Figure 12. Calculated morphology evolution from 1200 K to 1400 K with considering grain boundary constraint.

Table 1. Process parameters for Inconel 718 experiment [45].

| Parameter | Values | Parameter | Values |
|---|---|---|---|
| Laser power (W) | 285 | Laser diameter (μm) | 7 |
| Laser speed (mm/s) | 960 | Yt-fiber Laser Wavelength (nm) | 1070 |
| Hatch spacing (mm) | 0.11 | Spot Size (μm) | 70 to 80 |
| Scanning stripe width (mm) | 10 | Focal Length (mm) | 410 |
| Layer thickness (μm) | 40 | Time for one layer (s) | 40-60 |
| Recoater material | High speed steel | Time between layers | 13 |

Table 2. Phases predicted to show up (up to 1500 K) using the EDS compositional data for the 200 nm measurement as input.

| Phase | Primitive unit cell | Space group number | Point group symmetry | Space group symbol |
|---|---|---|---|---|
| AlNb2 | Al10Nb20 | 136 | D4h-14 | P4_2/mnm |
| AlNi2 | Al2Ni4 | 164 | D3d-3 | P-3m1 |
| C14 | CrNi2 | 71 | D2h-25 | Immm |
| C14 | MoNi2 | 71 | D2h-25 | Immm |
| Fe7Mo6 | Fe7Mo6 | 166 | D3d-5 | R-3m |
| AlNbNi2 | AlNbNi2 | 225 | Oh-5 | Fm-3m |
| AlNi3 | AlNi3 | 225 | Oh-5 | Fm-3m |
| L10 | FeNi | 123 | D4h-1 | Pm-3m |
| γ' | FeNi3 | 127 | D4h-5 | Pm-3m |
| γ' | CrFe3 | 127 | D4h-5 | Pm-3m |
| γ' | AlNi3 | 221 | Oh-1 | Pm-3m |
| M23C6 | C6Cr21Mo2 | 225 | Oh-5 | Fm-3m |
| NbFe2 | Fe8Nb4 | 194 | D6h-4 | P6_3/mmc |
| Ni4Mo | MoNi4 | 87 | C4h-5 | I4/m |
| bcc | Fe6Mo6Ni6 | 8 | Cs-3 | Cm |
| α-Cr | Cr127Mo | 221 | Oh-1 | Pm-3m |
| α-Fe | CrFe127 | 221 | Oh-1 | Pm-3m |
| bcc | Cr8Mo8 | 51 | D2h-5 | Pmma |
| bcc | Cr6Mo6Ni6 | 8 | Cs-3 | Cm |
| α-Cr | Cr | 221 | Oh-1 | Pm-3m |
| bcc | Mo | 229 | Oh-9 | Im-3m |
| bcc | Nb | 229 | Oh-9 | Im-3m |
| α-Cr | Cr127Fe | 221 | Oh-1 | Pm-3m |
| δ | Mo2Ni6 | 59 | D2h-13 | Pmmn |
| δ | Nb2Ni6 | 59 | D2h-13 | Pmmn |
| η | Ni12Ti4 | 194 | D6h-4 | P6_3/mmc |
| γ | Fe8Ni16 | 6 | Cs-1 | Pm |
| γ | Fe8Ni24 | 1 | C1-1 | P1 |
| γ | Fe5Mo5Ni15 | 1 | C1-1 | P1 |
| γ | Fe8Ni8 | 166 | D3d-5 | R-3m |
| γ | Fe24Ni40 | 1 | C1-1 | P1 |
| γ | Cr8Ni8 | 166 | D3d-5 | R-3m |
| γ | Co5Fe5Ni15 | 1 | C1-1 | P1 |
| γ | Al5Fe5Ni15 | 1 | C1-1 | P1 |
| γ | Co8Fe24 | 1 | C1-1 | P1 |
| γ | Cr5Fe15Ni5 | 1 | C1-1 | P1 |
| γ | Cr6Fe6Ni6 | 1 | C1-1 | P1 |
| γ | Cr8Ni16 | 6 | Cs-1 | Pm |
| γ | Cr5Mo5Ni15 | 1 | C1-1 | P1 |

Table 3. Phase fractions (%) predicted without considering grain boundary constraint using the EDS compositional data for the measurement at 200 nm scale as input.

| Phase | Primitive unit cell formula | 600 K | 800 K | 1000 K | 1200 K | 1400 K | X11 color names used in the plot |
|---|---|---|---|---|---|---|---|
| AlNb2 | Al10Nb20 | 0.10 | 0.10 | 0.10 | 0.10 | 0.10 | green |
| AlNi2 | Al2Ni4 | | | 0.10 | 0.19 | 0.33 | darkgreen |
| C14 | CrNi2 | 0.16 | 0.17 | | | | blueviolet |
| C14 | MoNi2 | 0.50 | 3.56 | 3.30 | 0.38 | | blueviolet |
| C14 | total= | 0.66 | 3.73 | 3.30 | 0.38 | | |
| Fe7Mo6 | Fe7Mo6 | 0.24 | 0.24 | 0.25 | 0.16 | | darkblue |
| AlNbNi2 | AlNbNi2 | 0.03 | 0.03 | 0.03 | 0.07 | 0.17 | gold |
| AlNi3 | AlNi3 | | | | | | gray |
| L10 | FeNi | 22.78 | | | | | cornflowerblue |
| γ' | FeNi3 | 27.40 | 7.86 | | | | chartreuse |
| γ' | CrFe3 | | | 0.08 | | | chartreuse |
| γ' | AlNi3 | 4.19 | 4.19 | 4.05 | 3.94 | 3.67 | chartreuse |
| γ' | total= | 31.59 | 12.05 | 4.13 | 3.94 | 3.67 | |
| M23C6 | C6Cr21Mo2 | 0.31 | 0.31 | 0.31 | 0.31 | 0.30 | grey |
| NbFe2 | Fe8Nb4 | 0.15 | 0.15 | 0.15 | 0.15 | 0.16 | magenta |
| Ni4Mo | MoNi4 | 0.11 | 0.12 | 1.60 | 1.52 | | firebrick |
| bcc | Fe6Mo6Ni6 | | | | | 0.05 | blue |
| α-Cr | Cr63Cr64Mo | | | 12.39 | 15.03 | 0.62 | blue |
| bcc | CrFe127 | 0.19 | 0.19 | | | | blue |
| bcc | Cr8Mo8 | | | | | 0.12 | blue |
| bcc | Cr6Mo6Ni6 | | | | | 0.16 | blue |
| α-Cr | CrCr | 20.89 | 20.89 | 8.52 | 0.10 | | blue |

|  |  | | | | |  |
|---|---|---|---|---|---|---|
| bcc | Mo | 0.02 | 0.02 | | 0.06 | 0.02 | blue |
| bcc | Nb | 0.12 | 0.12 | 0.11 | 0.11 | 0.10 | blue |
| bcc | Cr127Fe | | | | | | blue |
| bcc | total= | 21.22 | 21.21 | 21.04 | 15.30 | 1.08 | |
| δ | Mo2Ni6 | 5.78 | 1.65 | 0.42 | 0.38 | | darkolivegreen |
| δ | Nb2Ni6 | 12.12 | 12.12 | 12.10 | 12.06 | 11.93 | darkolivegreen |
| δ | total= | 17.89 | 13.77 | 12.53 | 12.44 | 11.93 | |
| η | Ni12Ti4 | 4.31 | 4.31 | 4.31 | 4.30 | 4.28 | cyan |
| γ | Fe8Ni16 | | 19.39 | 32.85 | 37.65 | 12.61 | deepskyblue |
| γ | Fe8Ni24 | | | 6.64 | 0.01 | 0.01 | deepskyblue |
| γ | Fe5Mo5Ni15 | | | | 4.89 | 0.80 | deepskyblue |
| γ | Fe8Ni8 | | 7.60 | 11.66 | 0.21 | 0.11 | deepskyblue |
| γ | Fe24Ni40 | | 16.39 | | | | deepskyblue |

| γ | Cr8Ni8 | | | | | deepskyblue |
|---|---|---|---|---|---|---|
| γ | Co5Fe5Ni15 | | 0.11 | 0.13 | 0.15 | 0.31 | deepskyblue |
| γ | Al5Fe5Ni15 | | | | | deepskyblue |
| γ | Co8Fe24 | 0.23 | 0.09 | | | deepskyblue |
| γ | Cr5Fe15Ni5 | | | 0.24 | 0.37 | 0.37 | deepskyblue |
| γ | Cr6Fe6Ni6 | | | | 14.58 | 42.43 | deepskyblue |
| γ | Cr8Ni16 | | | 0.17 | 2.85 | 13.78 | deepskyblue |
| γ | Cr5Mo5Ni15 | | | | | 7.17 | deepskyblue |
| γ | total= | 0.23 | 43.57 | 51.69 | 60.71 | 77.58 | |

Table 4. Phase fraction*s (%)* predicted with considering grain boundary constraint using the EDS compositional data for the measurement at 200 nm scale as input.

| Phase | Primitive unit cell formula | 600 K | 800 K | 1000 K | 1200 K | 1400 K | X11 color names used in the plot |
|---|---|---|---|---|---|---|---|
| C14 | CrNi2 | 0.16 | 0.17 | | | | blueviolet |
| C14 | MoNi2 | 0.06 | 0.38 | 0.35 | 0.04 | | blueviolet |
| C14 | total= | 0.23 | 0.55 | 0.35 | 0.04 | | |
| Cr23C6 | C6Cr23 | 0.02 | 0.09 | 0.10 | 0.26 | 0.04 | greenyellow |
| D022Nb | NbNi3 | 10.68 | 10.67 | 10.65 | 10.63 | 10.48 | darkcyan |
| Fe7Mo6 | Fe7Mo6 | 2.46 | 1.09 | 1.00 | 0.17 | | darkgreen |
| Fm-3m-AlNbNi2 | AlNbNi2 | | 0.03 | 0.03 | 0.07 | 0.21 | gray |
| L10 | FeNi | 15.28 | | | | | cornflowerblue |
| L12 | Ni3Ti | 3.77 | 3.79 | 3.79 | 3.77 | 3.77 | chartreuse |
| L12 | FeNi3 | 37.38 | 6.64 | | | | chartreuse |
| L12 | AlNi3 | 4.19 | 4.19 | 4.15 | 4.14 | 4.00 | chartreuse |
| L12 | total= | 45.33 | 14.61 | 7.94 | 7.92 | 7.78 | |
| NI-Ni2Mo4C | C4Mo16Ni8 | 0.40 | 0.29 | 0.27 | 0.03 | | gold |
| NbFe2 | Fe8Nb4 | 0.16 | 0.15 | 0.15 | 0.15 | 0.16 | magenta |
| Ni4Mo | MoNi4 | 0.62 | 2.39 | 2.53 | 2.03 | | firebrick |
| bcc | CrFe127 | 0.19 | 0.19 | | | | blue |
| bcc | Cr63Cr64Mo | | | 14.58 | 15.07 | 0.64 | blue |
| bcc | CrCr | 21.21 | 21.13 | 6.57 | 0.10 | | blue |
| bcc | Mo | 0.02 | 0.35 | 0.31 | 0.05 | 0.02 | blue |
| bcc | Nb | 0.19 | 0.19 | 0.18 | 0.18 | 0.18 | blue |
| bcc | Cr6Mo6Ni6 | | | | | 0.24 | blue |
| bcc | Cr8Mo8 | | | | | 0.10 | blue |
| bcc | total= | 21.61 | 21.87 | 21.63 | 15.40 | 1.19 | |
| delta | Mo2Ni6 | 0.59 | 0.16 | 0.04 | 0.04 | | darkolivegreen |
| delta | Nb2Ni6 | 1.45 | 1.45 | 1.44 | 1.44 | 1.42 | darkolivegreen |
| delta | total= | 2.04 | 1.61 | 1.48 | 1.48 | 1.42 | |
| eta | Ni12Ti4 | 0.53 | 0.53 | 0.53 | 0.52 | 0.52 | cyan |
| gamma | Co8Fe24 | 0.23 | 0.05 | | | | deepskyblue |
| gamma | Fe24Ni40 | | 11.05 | | | | deepskyblue |
| gamma | Fe5Mo5Ni15 | | | | 5.45 | 0.77 | deepskyblue |
| gamma | Fe8Ni8 | | 3.84 | 6.88 | 0.18 | 0.12 | deepskyblue |
| gamma | Fe8Ni24 | | | 6.52 | 0.01 | 0.01 | deepskyblue |
| gamma | Cr8Ni16 | | | 0.17 | 2.80 | 13.76 | deepskyblue |
| gamma | Fe8Ni16 | | 30.65 | 38.77 | 37.24 | 12.39 | deepskyblue |

| | | | | | | |
|---|---|---|---|---|---|---|
| gamma | Co5Fe5Ni15 | | 0.09 | 0.10 | 0.15 | 0.31 | deepskyblue |
| gamma | Cr5Mo5Ni15 | | | | | 7.22 | deepskyblue |
| gamma | Cr5Fe15Ni5 | | | 0.28 | 0.44 | 0.38 | deepskyblue |
| gamma | Cr6Fe6Ni6 | | | | 14.54 | 42.65 | deepskyblue |
| gamma | total= | 0.23 | 45.69 | 52.73 | 60.80 | 77.60 | |